\documentclass[12pt]{article}
\usepackage{graphicx,amsmath,amssymb}
\usepackage[noadjust]{cite}
\usepackage{fancyhdr}

\begin{document}

\title{Chapter 10. Optical Properties of Graphene in External Fields}
\author{Y. H. Chiu$^{1,a}$, Y. C. Ou$^{2,b}$, and M. F. Lin$^{2,c}$ \\
$^{1}${\small National Center for Theoretical Sciences, Taiwan}\\
$^{2}${\small Department of Physics, National Cheng Kung University, Tainan,
Taiwan}\\
{\small $^{a}$e-mail address: airegg.py90g@nctu.edu.tw}\\
{\small $^{b}$e-mail address: l2897113@mail.ncku.edu.tw}\\
{\small $^{c}$e-mail address: mflin@mail.ncku.edu.tw}}
\maketitle

\begin{abstract}
The generalized tight-binding model, with the exact diagonalization method,
is developed to investigate optical properties of graphene in five kinds of
external fields. The quite large Hamiltonian matrix is transferred into the
band-like one by the rearrangement of many basis functions; furthermore, the
spatial distributions of wave functions on distinct sublattices are utilized
to largely reduce the numerical computation time. The external fields have a
strong influence on the number, intensity, frequency and structure of absorption
peaks, and the selection rules. The optical spectra in a uniform magnetic
field exhibit plentiful symmetric absorption peaks and obey a specific
selection rule. However, there are many asymmetric peaks and extra selection
rules under the modulated electric field, the modulated magnetic field, the
composite electric and magnetic fields, and the composite magnetic fields.
\end{abstract}

\newpage

\section{Introduction}

Monolayer graphene (MG), constructed from a single layer of carbon atoms
densely packed in hexagonal lattice, was successfully produced by mechanical
exfoliation.\cite{cpc1,cpc2} This particular material offers an excellent
system for studying two-dimensional (2D) physical properties, such as the
quantum Hall effects,\cite%
{qhe1,qhe2,qhe3,qhe4,qhe5,qhe6,qhe7,qhe8,qhe9,prb3,prb4,prb25,prb30} and
these properties could be preliminarily comprehended by the energy
dispersion (or called energy band structure), which can directly reflect the
main features of electronic properties. In the low-energy region of $%
\left\vert E^{c,v}\right\vert \leq 1 $ eV, MG possesses isotropic linear
bands crossing at the $\mathbf{K}$ ($\mathbf{K}^{\prime }$) point and is
regarded as a 2D zero-gap semiconductor, where $c$ ($v$) indicates the
conduction (valence) bands.\cite{p145.47} The linear bands are symmetric
about the Fermi level ($E_{F}=0$) and become nonlinear and anisotropic with $%
\left\vert E^{c,v}\right\vert >1$ eV.\cite{p145.47} Most importantly, the
quasiparticles related to the linear bands can be described by a Dirac-like
Hamiltonian,\cite{prb22} which is associated with relativistic particles and
dominates the low-energy physical properties.\cite{prb30,p207.26,p207.27}
Such a special electronic structure has been verified by experimental
measurements.\cite{cpc2,prb24}

MG has become a potential candidate of nano-devices due to its exotic
electronic properties. Well understanding the behavior of MG under external
fields is useful for improving the characteristics of graphene-based
nano-devices. Five cases of external fields (see table below), which can be
experimentally produced,\cite%
{p145.45,p145.46,cpc28,cpc29,cpc30,cpc31,p148.28} are often applied to
investigate the physical properties of few-layer graphenes (FLGs). In the
presence of a uniform magnetic field (UM), the electronic states
corresponding to the linear bands change into Landau levels (LLs) which obey
a specific relationship $E^{c,v}\propto \sqrt{n^{c,v}B_{UM}}$, where $n^{c}$
($n^{v}$) is the quantum number of the conduction (valence) states and $%
B_{UM}$ the magnetic field strength. The related anomalous quantum Hall
effects and particular optical excitations have been verified experimentally.%
\cite{prb25,prb30} For a modulated magnetic field (MM), quasi-Landau levels
(QLLs) possessing anisotropic behavior and the related optical absorption
spectra with specified selection rules were shown.\cite{p145,p166}
Furthermore, Haldane predicted that MG in the modulated magnetic field could
reveal quantum Hall effects even without any net magnetic flux through the
whole space.\cite{prb4} Concerning a modulated electric potential (ME), the
linear dispersions become oscillatory and extra Dirac cones are induced by
the potential.\cite%
{electricfield1,electricfield2,electricfield3,p97,p151,p148} Such a
potential changes MG from a zero-gap semiconductor into a semimetal$\ $\cite%
{p97,p148} and makes MG exhibit Klein paradox effect associated with the
Dirac cones.\cite{prb22} For two cases of composite fields, a uniform
magnetic field combined with a modulated magnetic field (UM-MM) and a
uniform magnetic field combined with a modulated electric potential (UM-ME),
the LL properties are drastically changed by the modulated fields. For both
composite field cases, an unusual oscillation \cite%
{weiss1,weiss2,weiss3,weiss4,weiss5} of the density of states (DOS) similar
to the Weiss oscillation obtained in 2D electron gas (2DEG) were shown.
Furthermore, the broken symmetry, displacement of the center location, and
alteration of the amplitude strength of the LL wave functions were also
obtained.\cite{p197,cpc}

Graphene-related systems are predicted to exhibit rich optical absorption
spectra. The spectral intensity of MG is proportional to the frequency, but
no prominent peak exists at $\omega <5$ eV.\cite{p145.5} In FLGs, the
interlayer atomic interactions drastically alter the two linear energy bands
intersecting at $E_{F}=0$.\cite%
{p164.6,p164.7,p164.8,p164.9,p164.10,p164.11,p164.12,p164.13,p164.14,p96,p107}
As a result, conspicuous absorption peaks arise in optical spectra,\cite%
{p105} where the peak structure, intensity and frequency are dominated by
the layer number and the stacking configuration. Furthermore, under an
external perpendicular electric field or a uniform perpendicular magnetic
field, the main features of the optical properties of the FLGs are strongly
modified.\cite{p164.11,p164.30,RBChen} For theoretical studies, the
complexity of calculating the optical absorption spectra is solved by the
gradient approximation based on the generalized tight-binding model with
exact-diagonalization method or effective-mass approximation. The way in
which one can control the absorption peaks and selection rules is worthy to
be reviewed in detail.

On the other hand, there has been a considerable amount of experimental
research on graphene-related systems under a uniform perpendicular magnetic
field. From the measured results, the features of MG and bilayer graphene
are reflected in the magneto-optical spectra.\cite{prb165443,np621} That is
to say, the LL energies are proportional to $\sqrt{n^{c,v}B_{0}}$ or $%
n^{c,v}B_{0}$.\ For any graphene system, the selection rule coming from the
LLs close to $E_{F}=0$ is $\Delta n=|n^{c}-n^{v}|=1$.\cite%
{any1,bilayer1,abgraphite1} Moreover, similar results may also be found in
AB-stacked graphite.\cite{abgraphite1} However, experimental measurements on
optical properties under a non-uniform or composite fields are not available
so far.

In this chapter, we would like to focus on the optical absorption spectra of
monolayer graphene under the five cases of external fields, UM, MM, ME,
UM-MM and UM-ME cases. The tight-binding (TB) model with exact
diagonalization method is introduced to solve the energy dispersions and
then the gradient approximation is applied to obtain the optical absorption
spectra. The main features of electronic properties, which include energy
dispersions and wave functions, will be shown to comprehend the optical
absorption properties, where the dependence of absorption frequency on
external fields, optical selection rules and anisotropic behavior will be
discussed in detail. In Sec. 10.1, the tight-binding model corresponding to
the five cases of external fields is shown. In Secs. 10.2 to 10.6, the
optical absorption spectra of MG\ under the UM, MM, ME, UM-MM and UM-ME
cases will be reviewed, respectively. Finally, concluding remarks are
presented in Sec. 10.7.

\begin{tabular}{|p{3in}|p{3in}|}
\hline
\multicolumn{2}{|c|}{Physical properties of graphene under external fields}
\\ \hline
External fields & Related physical properties \\ \hline
Uniform magnetic field & Landau level and Abnormal quantum Hall effect,\cite%
{qhe1,qhe2,qhe3,qhe4,qhe5,qhe6,qhe7,qhe8,qhe9,prb3,prb4,prb25,prb30}
Magneto-optical selection rule\cite{prb165443,np621} \\ \hline
Modulated magnetic field & Quasi-Landau level,\cite{p145,p166} Quantum Hall
effect without Landau level\cite{prb4} \\ \hline
Modulated electric potential & Number increasement of Dirac cone,\cite%
{electricfield1,electricfield2,electricfield3,p97,p151,p148}
Semiconductor-metal transition,\cite{p97,p148} Klein tunneling\cite{prb22}
\\ \hline
Uniform magnetic field+Modulated magnetic field, Uniform magnetic
field+Modulated electric potential & Weiss oscillation,\cite%
{weiss1,weiss2,weiss3,weiss4,weiss5} Destruction of Landau-level wavefunction%
\cite{p197,cpc} \\ \hline
\end{tabular}

\newpage

\section{Tight-Binding Model with Exact Diagonalization}

The low-frequency optical properties of graphene are determined by the $\pi $%
-electronic structure due to the 2$p_{z}$ orbitals of carbon atoms. The
generalized tight-binding model with exact diagonalization method is
developed to characterize the electronic properties and then the gradient
approximation is applied to obtain the optical-absorption spectra. In the
absence of external fields, there are two carbon atoms, the $a$ and $b$
atoms, in a primitive unit cell of MG, as shown in Fig. 10.1(a) by the green
shadow, where the $x$- and $y$-direction are respectively the armchair and
zigzag directions of MG. This indicates that the Bloch wave function $\Psi $%
\ is a linear\ superposition of two TB functions associated with the 2$p_{z}$
orbitals and expressed as $\Psi =\varphi _{a}\pm \varphi _{b}$, where $%
\varphi _{a}$ and $\varphi _{b}$ respectively stand for the tight-binding
functions of the $a$ and $b$ atoms and are represented as \cite{p145.47}
\begin{align}
\varphi _{a} =\sum\limits_{a}\exp (i\mathbf{k}\cdot \mathbf{R}_{a})\chi (%
\mathbf{r}-\mathbf{R}_{a})\text{,}  \tag{10.1a} \\
\varphi _{b} =\sum\limits_{b}\exp (i\mathbf{k}\cdot \mathbf{R}_{b})\chi (%
\mathbf{r}-\mathbf{R}_{b})\text{.}  \tag{10.1b}
\end{align}%
$\chi (\mathbf{r})$ is the normalized orbital 2$p_{z}$ wave function for an
isolated atom. Moreover, the symbols $\gamma _{0}$ ($=2.5$ eV) and $%
b^{\prime }$ ($=1.42$ \AA ) shown in Fig. 10.1(a) represent the
nearest-neighbor atomic interaction (or called hopping integral) and the C-C
bond length, respectively.\cite{p145.47} Throughout this chapter, only $%
\gamma _{0}$ is taken into account and other atomic interactions are
neglected.

\begin{figure}[tbp]
\par
\begin{center}
\leavevmode
\includegraphics[width=0.8\linewidth]{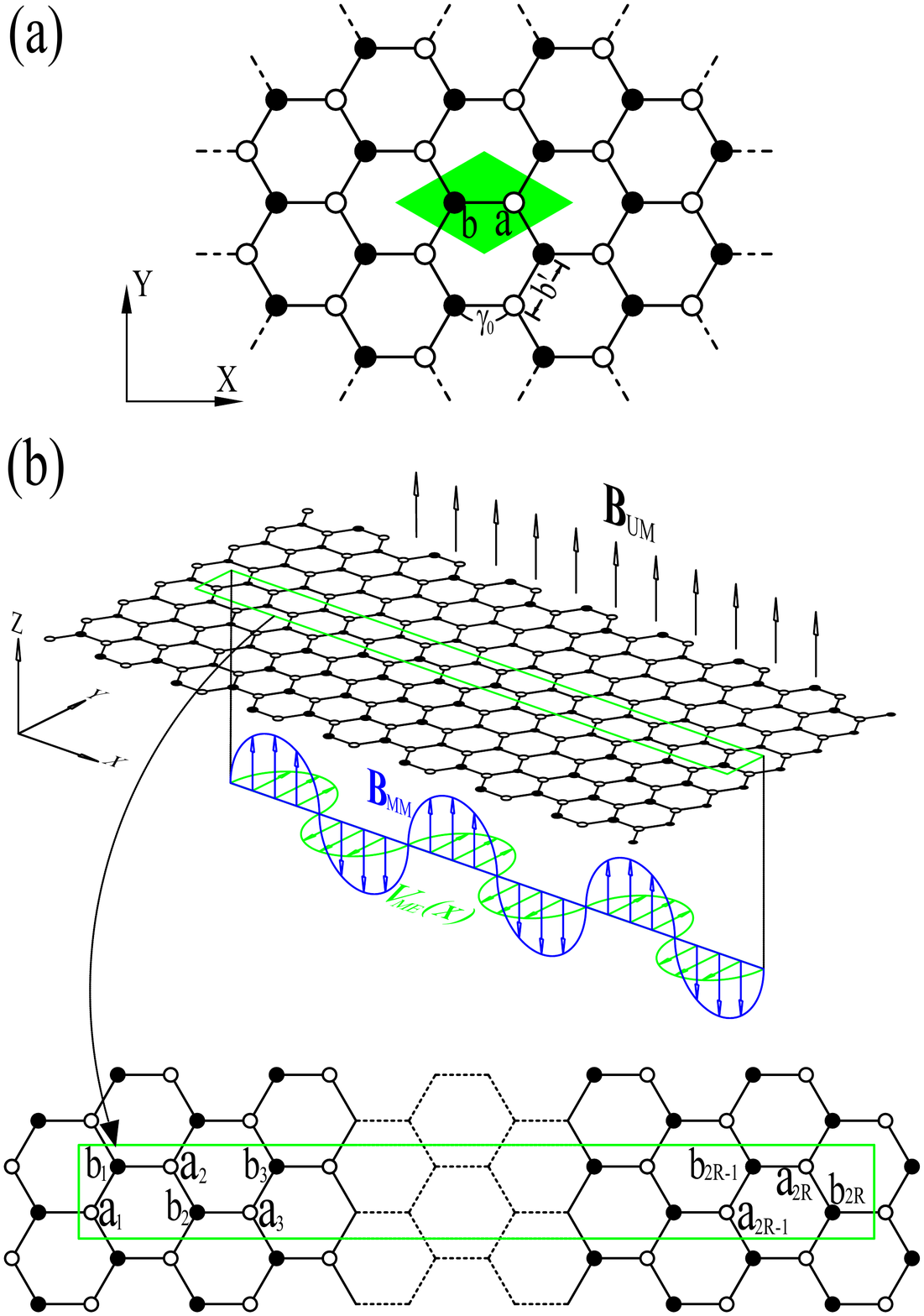}
\end{center}
\par
\textbf{Figure 10.1.} The primitive unit cell of monolayer graphene (a) in the absence
and (b) in the presence of external fields.
\end{figure}

In the presence of an external field, the primitive unit cell is no longer
the one shown in Fig. 10.1(a) since the external field leads to a new
periodic condition. Here we choose the rectangular unit cell marked by the
green rectangle in Fig. 10.1(b) as the primitive unit cell of graphene under
the five kinds of external fields, where $R=R_{UM}$, $R_{MM}$, $R_{ME}$, and
$R_{C}$ (defined in the following) describe respectively the periods
resulting from the uniform magnetic field, modulated magnetic field,
modulated electric potential, and composite field. The major discussions are
focused on $R$ along the armchair direction. Consequently, an enlarged
rectangular unit cell induced by an external field encompasses $2R$ $a$
atoms and $2R$ $b$ atoms. This implies that $R$ determines the dimension of
the Hamiltonian matrix, which is a $4R\times 4R$ Hermitian\ matrix spanned
by $4R$ TB functions associated with the $2R$ $a$ atoms and $2R$ $b$ atoms.
Based on the arrangement of odd and even atoms in the primitive cell, the
Bloch wave function $|\Psi _{\mathbf{k}}\rangle $\ can have the expression:
\begin{equation}
|\Psi _{\mathbf{k}}\rangle =\sum\limits_{m=1}^{2R-1}(A_{\mathbf{o}%
}^{c,v}|a_{m\mathbf{k}}\rangle +B_{\mathbf{o}}^{c,v}|b_{m\mathbf{k}}\rangle
)+\sum\limits_{m=2}^{2R}(A_{\mathbf{e}}^{c,v}|a_{m\mathbf{k}}\rangle +B_{%
\mathbf{e}}^{c,v}|b_{m\mathbf{k}}\rangle )\text{.}  \tag{10.2}
\end{equation}%
$|a_{m\mathbf{k}}\rangle $ ($|b_{m\mathbf{k}}\rangle $) is the TB function
corresponding to\ the 2$p_{z}$ orbital of the $m$th $a$ ($b$) atom. $A_{%
\mathbf{o}}^{c,v}$ ($A_{\mathbf{e}}^{c,v}$) and $B_{\mathbf{o}}^{c,v}$ ($B_{%
\mathbf{e}}^{c,v}$) are the subenvelope functions standing for the
amplitudes of the wave functions of the $a$- and $b$-atoms respectively,
where $o$ ($e$) represents an odd (even) integer. Since the features of $%
A_{o}^{c,v}$ ($B_{o}^{c,v}$) and $A_{e}^{c,v}$ ($B_{e}^{c,v}$)\ are similar,
choosing only the amplitudes $A_{\mathbf{o}}^{c,v}$ and $B_{\mathbf{o}%
}^{c,v} $ is sufficient to comprehend the electronic and optical properties
we would like to discuss in this chapter. The $4R\times 4R$ Hamiltonian
matrix, which determines the magneto-electronic properties, is a giant
Hermitian matrix for the external fields actually used in experiments. To
make the calculations more efficient, the matrix is transformed into an $%
M\times 4R$ band-like matrix by a suitable rearrangement of the
tight-binding functions, where $M$ is much smaller than $4R$. For example,
one can arrange the basis functions as the sequence: $|a_{1\mathbf{k}%
}\rangle $, $|b_{2R\mathbf{k}}\rangle $, $|b_{1\mathbf{k}}\rangle $, $|a_{2R%
\mathbf{k}}\rangle $, $|a_{2\mathbf{k}}\rangle $, $|b_{2R-1\mathbf{k}%
}\rangle $, $|b_{2\mathbf{k}}\rangle $, $|a_{2R-1\mathbf{k}}\rangle $, ......%
$|a_{R-1\mathbf{k}}\rangle $, $|b_{R+2\mathbf{k}}\rangle $, $|b_{R-1\mathbf{k%
}}\rangle $, $|a_{R+2\mathbf{k}}\rangle $, $|a_{R\mathbf{k}}\rangle $, $%
|b_{R+1\mathbf{k}}\rangle $, $|b_{R\mathbf{k}}\rangle $; $|a_{R+1\mathbf{k}%
}\rangle $. Furthermore, distributions of the subenvelope functions are used
to reduce the numerical computation time. The exact diagonalization method
for numerical calculations is applicable to many kinds of magnetic, electric
and composite fields.

For the UM case $\mathbf{B}_{UM}=B_{UM}\widehat{z}$, a\ Peierls phase \cite%
{p145,p121,peierls2,peierls3} related to the vector potential $\mathbf{A}%
_{UM}\mathbf{=}B_{UM}x\widehat{y}$ is introduced in the TB functions. The
phase difference between two lattice vectors\ ($\mathbf{R}_{m}$ and $\mathbf{%
R}_{m^{\prime }}$) is\ defined as $G_{UM}\equiv {\frac{2{\pi }}{{\phi }_{0}}}
$ $\int_{\mathbf{R}_{m^{\prime }}}^{\mathbf{R}_{m}}\mathbf{A}_{UM}\cdot d%
\mathbf{r}$ , where $\phi _{0}=hc/e=4.1356\times 10^{-15}$ $[$T m$^{2}]$ is
the flux quantum. The Peierls phase periodic along the armchair direction
provides a specific period set as $R_{UM}=\frac{\phi _{0}/(3\sqrt{3}%
b^{\prime 2}/2)}{B_{UM}}$ and the related Hamiltonian is a $4R_{UM}\times
4R_{UM}$ Hermitian\ matrix. The site energies, the diagonal matrix elements $%
\langle a_{m\mathbf{k}}|H|a_{m\mathbf{k}}\rangle $ and $\langle b_{m\mathbf{k%
}}|H|b_{m\mathbf{k}}\rangle $,\ are\ set to zero and the nonzero\ matrix\
elements related to $\gamma _{0}$\ can be formulated as%
\begin{equation}
\langle b_{m\mathbf{k}}|H|a_{m^{\prime }\mathbf{k}}\rangle =\gamma _{0}\exp
i[\mathbf{k\cdot (\mathbf{R}}_{m}\mathbf{-\mathbf{R}}_{m^{\prime }}\mathbf{)}%
+{G}_{UM}]\text{.}  \tag{10.3}
\end{equation}

Two kinds of periodic modulation fields along the armchair direction, the MM
and ME cases, which can drastically change the physical properties of MG,
are often selected for a study. For the MM case, $\mathbf{B}_{MM}=B_{MM}\sin
(2\pi x/l_{MM})$\ $\widehat{z}$ is exerted on\ MG along the armchair
direction, where $B_{MM}$ is the field strength and $l_{MM}$ is the period
length with the modulation period $R_{MM}=l_{MM}/3b^{\prime }$. The vector
potential is chosen as $\mathbf{A}_{MM}\mathbf{=(-}B_{MM}\frac{l_{B}}{2\pi }%
\cos (2\pi x/l_{MM})\mathbf{)}\widehat{y}$ and the corresponding Peirls
phase is $G_{MM}\equiv {\frac{2{\pi }}{{\phi }_{0}}}$ $\int_{\mathbf{R}%
_{m^{\prime }}}^{\mathbf{R}_{m}}\mathbf{A}_{MM}\cdot d\mathbf{r}$. Thus the
Hamiltonian\ matrix elements, which are similar to those in Eq. (10.3), are
represented as
\begin{equation}
\langle b_{m\mathbf{k}}|H|a_{m^{\prime }\mathbf{k}}\rangle =\gamma _{0}\exp
i[\mathbf{k\cdot (\mathbf{\mathbf{R}}}_{m}\mathbf{\mathbf{-\mathbf{R}}}%
_{m^{\prime }}\mathbf{)}+{G}_{MM}]\text{.}  \tag{10.4}
\end{equation}%
For the ME case, $V_{ME}(x)=V_{ME}\cos (2\pi x/l_{ME})$ along the armchair
direction with the potential strength $V_{ME}$ and the period length $l_{ME}$
is taken into account. As the period is sufficiently large, the electric
potential affects only the site energies but not the nearest-neighbor
hopping integral. As a result, the site energies become
\begin{subequations}
\begin{align}
\langle a_{m\mathbf{k}}|H|a_{m\mathbf{k}}\rangle & =V_{ME}\cos [(m-1)\pi
/R_{ME}]\equiv V_{m}\text{,}  \tag{10.5a} \\
\langle b_{m\mathbf{k}}|H|b_{m\mathbf{k}}\rangle & =V_{ME}\cos [(m-2/3)\pi
/R_{ME}]\equiv V_{m+1/3}\text{,}  \tag{10.5b}
\end{align}%
where $R_{ME}=l_{ME}/3b^{\prime }$ is the modulation period. The Hamiltonian
matrices for the modulated magnetic field and the modulated electric
potential are $4R_{MM}\times 4R_{MM}$\ and$\ 4R_{ME}\times 4R_{ME}$
Hermitian\ matrices, respectively.

For a composite field case, a new periodicity, which is associated with
periods induced by a uniform magnetic field and a modulated field, has to be
defined. The rectangular unit cell is enlarged along the $x$-direction and
the dimensionality of the Hamiltonian matrix has to agree with the least
common multiple of $R_{UM}$ and $R_{MM}$ ($R_{UM}$ and $R_{ME}$) for the
UM-MM (UM-ME) case, namely $R_{C}$. The rectangular unit cell corresponding
to each composite field contains $4R_{C}$ atoms\ ($2R_{C}$ $a$ atoms and $%
2R_{C}$ $b$ atoms), and the magneto-electronic wave functions are linear
combinations of the $4R_{C}$ TB functions. In a composite field case, the
matrix elements are superposed by the elements associated with each combined
external field. For the sake of convenience, we put the matrix elements in
Eqs. (10.3)-(10.5) together as a common case and the elements are rewritten
as
\end{subequations}
\begin{subequations}
\begin{align}
\langle b_{m\mathbf{k}}|H|a_{m^{\prime }\mathbf{k}}\rangle & =\gamma
_{0}\exp i[\mathbf{k\cdot (\mathbf{R}}_{m}\mathbf{-\mathbf{R}}_{m^{\prime }}%
\mathbf{)}+{G}_{UM}+G_{MM}]\text{,}  \tag{10.6a} \\
\langle a_{m\mathbf{k}}|H|a_{m\mathbf{k}}\rangle & =V_{m}\text{,}
\tag{10.6b} \\
\langle b_{m\mathbf{k}}|H|b_{m\mathbf{k}}\rangle & =V_{m+1/3}\text{.}
\tag{10.6c}
\end{align}%
The off-diagonal elements are associated with the Peierls phases induced by
the magnetic fields and the diagonal elements are related to the site
energies induced by the modulated electric field. By diagonalizing the
matrix, the energy dispersion $E^{c,v}$ and the wave function $\Psi ^{c,v}$
are obtained. It should be noted that the $k_{x}$-dependent dispersions can
be ignored when the period $R$ is sufficiently large and thus only $k_{y}$%
-dependent dispersions are shown for the following discussions.

When a monolayer graphene is excited from the occupied valence to the
unoccupied conduction bands by an electromagnetic field,\ only inter-$\pi $%
-band excitations exist at zero temperature. Based on the Fermi's golden
rule, the optical absorption function results in the following form
\end{subequations}
\begin{align}
A(\omega ) \propto &\sum\limits_{c,v,\widetilde{n},\widetilde{n}^{\prime
}}\int_{1stBZ}\frac{d\mathbf{k}}{(2\pi )^{2}}\left\vert \langle \Psi ^{c}(%
\mathbf{k},n)|\frac{\widehat{\mathbf{E}}\cdot \mathbf{P}}{m_{e}}|\Psi ^{v}(%
\mathbf{k},n^{\prime })\rangle \right\vert ^{2}  \notag \\
&\times \mathrm{Im}\left[ \frac{f(E^{c}(\mathbf{k},n))-f(E^{v}(\mathbf{k}%
,n^{\prime }))}{E^{c}(\mathbf{k},n)-E^{v}(\mathbf{k},n^{\prime })-\omega
-i\Gamma }\right] \text{,}  \tag{10.7}
\end{align}%
where $f(E(\mathbf{k},\widetilde{n}))$ is the Fermi-Dirac distribution
function, and $\Gamma $ ($=2\times \,10^{-4}\gamma _{0}$) is the broadening
parameter. The electric polarization $\widehat{\mathbf{E}}$ is the unit
vector of an electric polarization. Results for $\widehat{\mathbf{E}}$ along
the armchair and zigzag directions are taken into account for discussions.
Within the gradient approximation,\cite{cpc37,cpc39,cpc40} the velocity
matrix element $M^{cv}=\langle \Psi ^{c}(\mathbf{k},\widetilde{n})|\frac{%
\widehat{\mathbf{E}}\cdot \mathbf{P}}{m_{e}}|\Psi ^{v}(\mathbf{k},\widetilde{%
n}^{\prime })\rangle $ is formulated as%
\begin{equation}
\sum\limits_{m,m\prime =1}^{2R_{C}}(A^{c}{}^{\ast }\times B^{v})\nabla _{%
\mathbf{k}}\langle a_{m\mathbf{k}}|H|b_{m^{\prime }\mathbf{k}}\rangle +h.c.%
\text{.}  \tag{10.8}
\end{equation}%
Equation (10.8) implies that the main features of the wave functions are
major factors in determining the selection rules and the absorption rate of
the optical excitations. Similar gradient approximations have been
successfully applied to explain optical spectra of carbon-related systems,
e.g., graphite,\cite{10.1.1} graphite intercalation compounds,\cite{10.1.2}
carbon nanotubes,\cite{10.1.3} few-layer graphenes,\cite{10.1.4} and
graphene nanoribbons.\cite{p145.46}

\newpage

\section{Uniform Magnetic Field}

\subsection{Landau Level Spectra}

In this section, we mainly focus on drastic changes of the Dirac cone as the
result of a uniform perpendicular magnetic field. The magnetic field causes
the states to congregate and induces dispersionless Landau levels, as shown
in Fig. 10.2(a) for\ $B_{UM}=5$ T at $k_{x}=0$. The unoccupied LLs and
occupied LLs\ are symmetric about the Fermi level ($E_{F}=0$). Each LL is
characterized by the quantum number\ $n^{c,v}$, which corresponds to the
the\ number of zeros in the eigenvectors of harmonic oscillator.\cite%
{p164.12,p179} Each LL is fourfold degenerate without considering the spin
degeneracy. Its energy may be approximated by a simple square-root
relationship $|E_{n}^{c,v}|\propto \sqrt{n^{c,v}B_{UM}}$,\cite{prb3,10.2.4}
which is valid only for the range of $|E_{n}^{c,v}|\leq \pm 1$ eV.\cite{prb3}

\begin{figure}[tbp]
\par
\begin{center}
\leavevmode
\includegraphics[width=0.8\linewidth]{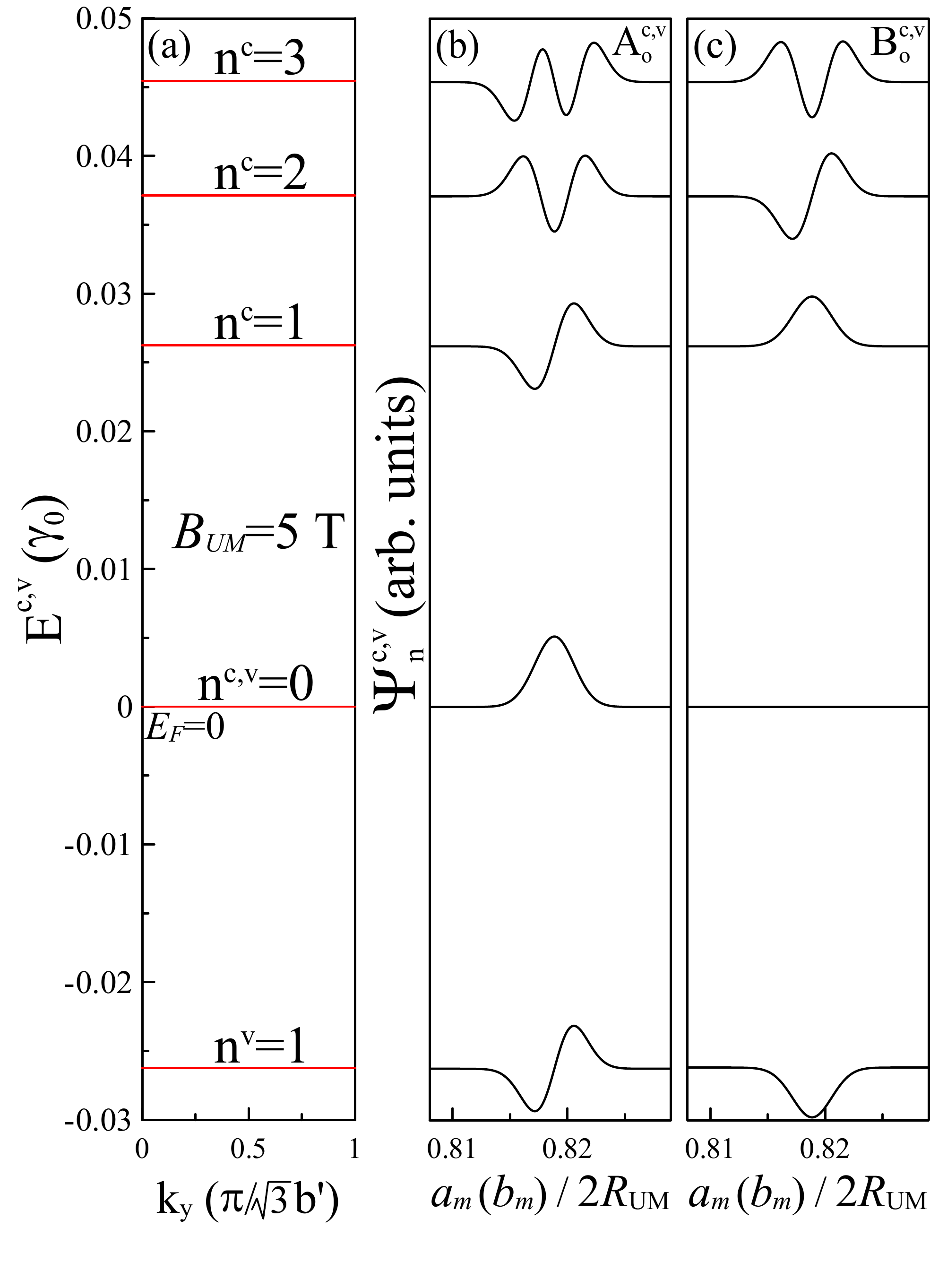}
\end{center}
\par
\textbf{Figure 10.2.} The Landau level spectrum for the uniform magnetic field $%
B_{UM}=5$ T. The Landau level wave functions corresponding to (b) the $a$-
and (c) the $b$-atoms.
\end{figure}

\subsection{Landau Level Wave Functions}

The LL wave functions, as shown in Figs. 10.2(b) and 10.2(c), exhibit the
versatility of spatial symmetry and can be described by the eigenvectors ($%
\varphi _{n}\left( x\right) $) of harmonic oscillator, which obey the
relationships, $\langle \varphi _{n}\left( x\right) |\varphi _{n^{\prime
}}\left( x\right) \rangle =\delta _{n,n\prime }$ and $\varphi _{n}\left(
x\right) =0$ for $n<0$. The wave functions are distributed around the
localization center, that is at the $5/6$ position of the enlarged unit
cell. Similar localization centers corresponding to the other degenerate
states occur at the $1/6$, $2/6$, and $4/6$ positions. The subenvelope
functions can be expressed as
\begin{subequations}
\begin{align}
A_{o,e}^{c,v}& \propto \varphi _{n^{c,v}}\left( x_{1}\right) \pm \varphi
_{n^{c,v}-1}\left( x_{2}\right) \text{, }B_{o,e}^{c,v}\propto \varphi
_{n^{c,v}-1}\left( x_{1}\right) \mp \varphi _{n^{c,v}}\left( x_{2}\right)
\text{,}  \notag \\
A_{o,e}^{c,v}& \propto \varphi _{n^{c,v}-1}\left( x_{3}\right) \pm \varphi
_{n^{c,v}}\left( x_{4}\right) \text{, }B_{o,e}^{c,v}\propto \varphi
_{n^{c,v}}\left( x_{3}\right) \mp \varphi _{n^{c,v}-1}\left( x_{4}\right)
\text{,}  \notag \\
\text{for }x_{1}& =1/6\text{, }x_{2}=5/6\text{, }x_{3}=2/6\text{, and }%
x_{4}=4/6\text{.}  \tag{10.9}
\end{align}%
However, it is adequate to only consider any one center in evaluating the
absorption spectra due to their identical optical responses.

\begin{figure}[tbp]
\par
\begin{center}
\leavevmode
\includegraphics[width=0.8\linewidth]{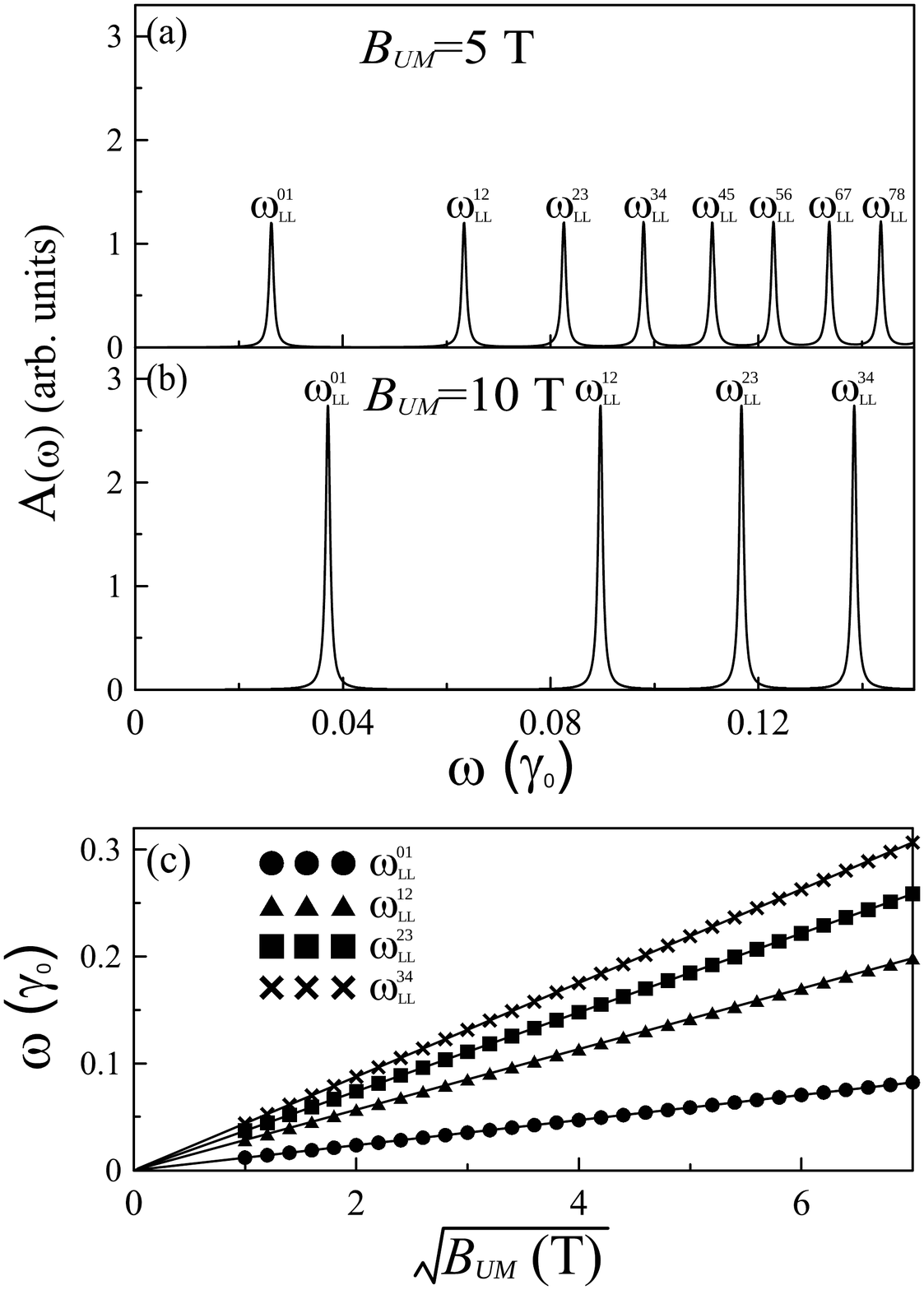}
\end{center}
\par
\textbf{Figure 10.3.} The optical absorption spectra for (a) $B_{UM}=5$ T and (b) $%
B_{UM}=10$ T. (c) The dependence of the absorption frequency on the square
root of field strength $B_{UM}$.
\end{figure}

\subsection{Optical Absorption Spectra of Landau Levels}

The low-frequency optical absorption spectrum of the LLs presents many
interesting features as shown in Fig. 10.3(a) for $B_{UM}=5$ T. The spectrum
exhibits many delta-function-like symmetric peaks with a uniform intensity.
Such peaks suggest that LLs possess a zero-dimensional (0D) band structure
or density of states. The optical transition channel with respect to each
absorption peak can be clearly identified. A single peak $\omega
_{LL}^{nn^{\prime }}$ is generated by two transition channels $n^{\prime }$LL%
$\rightarrow n$LL and $n$LL$\rightarrow n^{\prime }$LL, where the symbol $%
n^{\prime }\rightarrow n$ is used, for the sake of convenience, to represent
the transition from the valence states with $n^{\prime }$ to the conduction
states with $n$ throughout this chapter. The quantum numbers related to the
LL transitions must satisfy a specific selection rule, i.e., $\Delta
n=|n^{c}-n^{v}|=1$. The selection rule is established by the main features
of the wave functions. The velocity matrix $M^{cv}$, a dominant factor for
the excitations of the prominent peaks, strongly depends on the number of
zeros of $A_{\mathbf{o}}^{c,v}$ and $B_{\mathbf{o}}^{c,v}$. It has non-zero
values only when $A_{\mathbf{o}}^{c(v)}$ and $B_{\mathbf{o}}^{v(c)}$,
expressed in orthogonality of $\varphi _{n}\left( x\right) $, possess the
same number of zeros. Moreover, examining all the transitions reveals the
following relationship: $A_{\mathbf{o}}^{c,v}(n^{c,v})$ $\propto B_{\mathbf{o%
}}^{c,v}(n^{c,v}+1)$, with $A_{\mathbf{o}}^{c}=A_{\mathbf{o}}^{v}$ and $B_{%
\mathbf{o}}^{c}=-B_{\mathbf{o}}^{v}$. In other words, the quantum numbers of
the conduction and valence LLs differ by one when $A_{\mathbf{o}}^{c(v)}$
and $B_{\mathbf{o}}^{v(c)}$ have the same $\varphi _{n}\left( x\right) $.

In addition to the optical selection rules, the peak intensity and
absorption frequency also deserve a discussion. In Fig. 10.3(b), the peak
intensity is strengthened, whereas the peak number is reduced as the field
strength increases. This is a result of the high degree of degeneracy in the
first Brillouin zone and the expanded energy spacing between the LLs. The
field-dependent absorption frequencies of the first four peaks $\omega
_{LL}^{01}$, $\omega _{LL}^{12}$, $\omega _{LL}^{23}$, and $\omega
_{LL}^{34} $ are shown in Fig. 10.3(c). The frequencies become much higher
in a stronger field. There exists a special square-root relation between $%
\omega _{LL}^{nn^{\prime }}$ and $B_{0}$, i.e., $\omega _{LL}^{nn^{\prime
}}\propto \sqrt{B_{UM}}$, which has been confirmed by magneto-optical
spectroscopy methods, such as experimental measurements of the absorption
coefficient,\cite{10.2.6,10.2.7} cyclotron resonance,\cite%
{10.2.9,10.2.11,10.2.12} and quantum Hall conductivity.\cite%
{cpc2,10.2.13,10.2.16,10.2.17} This square-root relation only exists in the
lower frequency range $\omega <0.4\gamma _{0}$ ($\symbol{126}$1 eV). In the
higher frequency range, LLs are too densely packed to be separated from one
another.\cite{prb3} This leads to the disappearance of the relation between $%
\omega _{LL}^{nn^{\prime }}$\ and $B_{UM}$.

\newpage

\section{Spatially Modulated Magnetic Field}

\subsection{Quasi-Landau Level Spectra}

Compared with the uniform case, a modulated magnetic field has a different
impact on the electronic properties and leads to the diverse features
observed in the optical absorption spectra.\ The presence of a modulated
field has multiple effects on the energy bands, as shown in Fig. 10.4 for $%
B_{MM}=10$ T and $R_{MM}=500$. In the lower energy region, parabolic
subbands appear around $k_{y}=k_{1}=2/3$. The conduction and valence
subbands are symmetric about the Fermi level ($E_{F}=0$). The subbands
nearest to $E_{F}=0$ are partially flat and nondegenerate. The other
parabolic subbands characterized by weak energy dispersions have double
degeneracy and one original band-edge state at $k_{1}$. The modulation
effects on parabolic energy subbands result in four extra band-edge states
at the sites on both sides of $k_{1}$. They demonstrate the strongest
dispersion and destruction of the double degeneracy. The low-energy subbands
are regarded as quasi-Landau levels, which exhibit similar features of the
LLs generated from a uniform magnetic field. Moreover, the $k_{y}$ range
with respect to the weak dispersion and partial flat bands grows with
increasing field strength and a longer\ modulation period. On the contrary,
when the influence of the modulation field become much weaker with
increasing energy, the parabolic subbands in the higher energy region are
similar to the twofold degenerate subbands directly obtained from the zone
folding of MG in the $B_{MM}=0$ case (not shown).

\begin{figure}[tbp]
\par
\begin{center}
\leavevmode
\includegraphics[width=0.8\linewidth]{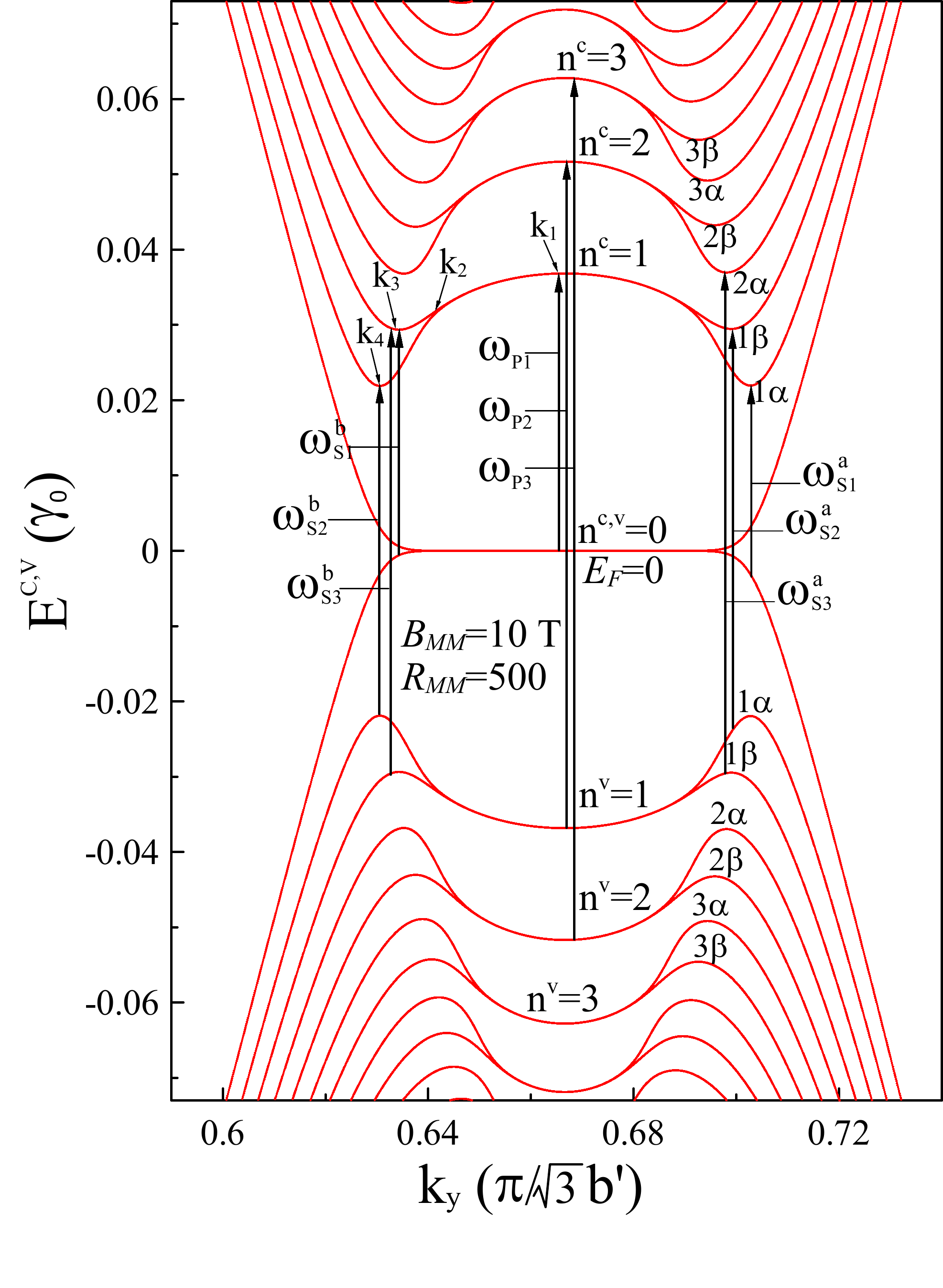}
\end{center}
\par
\textbf{Figure 10.4.} The energy dispersions and the illustration of optical excitation
channels for the modulated magnetic field along the armchair direction with $%
R_{MM}=500$ and $B_{MM}=10$ T.
\end{figure}

\subsection{Quasi-Landau Level Wave Functions}

In the presence of a modulated magnetic field, the alterations of the wave
functions are rather drastic. First, the QLL wave functions corresponding to
$k_{1}$ are shown in Figs. 10.5(a)-(f). The wave functions are composed of
two tight-binding functions centered at $x_{1}$ and $x_{2}$. $A_{\mathbf{o}%
}^{c}$ ($B_{\mathbf{o}}^{c}$) has two subenvelope functions $A_{\mathbf{o}%
}^{c}$($x_{1}$) ($B_{\mathbf{o}}^{c}$($x_{1}$)) and $A_{\mathbf{o}}^{c}$($%
x_{2}$) ($B_{\mathbf{o}}^{c}$($x_{2}$)) centered at $x_{1}=1/4$ and $%
x_{2}=3/4$ of the primitive unit cell, respectively. The positions $x_{1}$
and $x_{2}$ are located at where the field strength is at a maximum. The
number of zeros of $A_{\mathbf{o}}^{c}$($x_{2}$) ($B_{\mathbf{o}}^{c}$($%
x_{1} $)) is higher than that of $A_{\mathbf{o}}^{c}$($x_{1}$) ($B_{\mathbf{o%
}}^{c} $($x_{2}$)) by one at each QLL. A similar behavior is also shown by
the valence wave function, where only the sign is flipped in either $A_{%
\mathbf{o}}^{v}$ or $B_{\mathbf{o}}^{v}$. The effective quantum number $%
n^{c,v}$ is defined by the larger number of zeros of the subenvelope
functions. In addition, the twofold degenerate QLLs have similar wave
functions (black curves and red dashed curves), with the only difference in
terms of the sign change in the subenvelope functions. The wave functions at
$k_{1}$ can be expressed as
\end{subequations}
\begin{subequations}
\begin{align}
A_{o,e}^{c}& \propto \Psi _{n^{c}-1}\left( x_{1}\right) \pm \Psi
_{n^{c}}\left( x_{2}\right) \text{, }A_{o,e}^{v}\propto \Psi
_{n^{v}-1}\left( x_{1}\right) \mp \Psi _{n^{v}}\left( x_{2}\right) \text{,}
\notag \\
B_{o,e}^{c}& \propto \Psi _{n^{c}}\left( x_{1}\right) \mp \Psi
_{n^{c}-1}\left( x_{2}\right) \text{, }B_{o,e}^{v}\propto \Psi
_{n^{v}}\left( x_{1}\right) \pm \Psi _{n^{v}-1}\left( x_{2}\right) \text{,}
\notag \\
\text{for }x_{1}& =1/4\text{ and }x_{2}=3/4\text{.}  \tag{10.10}
\end{align}%
The wave functions would be strongly modified as the wave vectors gradually
move away from $k_{1}$. Secondly, the wave functions at several special $k$
points are illustrated to examine the effects caused by the modulated
magnetic field. As the wave vector moves to $k_{y}=k_{2}$, the doubly
degenerate QLL starts to separate into two subbands. The two subenvelope
functions $A_{\mathbf{o}}^{c}$($x_{1}$) ($B_{\mathbf{o}}^{c}$($x_{1}$)) and $%
A_{\mathbf{o}}^{c}$($x_{2}$) ($B_{\mathbf{o}}^{c}$($x_{2}$)) move toward
each other and shift to the center of the primitive unit cell with nearly
overlapping, as shown in Fig. 10.5(g) and 10.5(h). At $k_{3}$ and $k_{4}$,
the higher and lower subbands have the extra band-edge states $1\alpha $ and
$1\beta $, respectively. The subenvelope functions of the $1\alpha $ state,
as shown in Fig. 10.5(i) and 10.5(j), exhibit a strong overlapping behavior
compared to those at $k_{y}=k_{2}$ (red-dashed curves in Fig. 10.5(g) and
10.5(h)). Similar behavior can also be found in the wave functions at $%
1\beta $. This implies that there is a higher degree of overlap in the
subenvelope functions at the extra band-edge states $n^{c,v}\alpha $ and $%
n^{c,v}\beta $. Moreover, the two states associated with the different
linear combinations of $A_{\mathbf{o}}^{c}$($x_{1}$) ($B_{\mathbf{o}}^{c}$($%
x_{1}$)) and $A_{\mathbf{o}}^{c}$($x_{2}$) ($B_{\mathbf{o}}^{c}$($x_{2}$))
are represented as
\end{subequations}
\begin{subequations}
\begin{align}
A_{o,e}^{c}& \propto \Psi _{n^{c}-1}\left( x_{1}\right) +\Psi _{n^{c}}\left(
x_{2}\right) \text{ for }n^{c}\alpha \text{ and }\Psi _{n^{c}-1}\left(
x_{1}\right) -\Psi _{n^{c}}\left( x_{2}\right) \text{ for }n^{c}\beta \text{,%
}  \notag \\
B_{o,e}^{c}& \propto \Psi _{n^{c}}\left( x_{1}\right) -\Psi _{n^{c}-1}\left(
x_{2}\right) \text{ for }n^{c}\alpha \text{ and }\Psi _{n^{c}}\left(
x_{1}\right) +\Psi _{n^{c}-1}\left( x_{2}\right) \text{ for }n^{c}\beta
\text{,}  \notag \\
A_{o,e}^{v}& \propto \Psi _{n^{v}-1}\left( x_{1}\right) -\Psi _{n^{v}}\left(
x_{2}\right) \text{ for }n^{v}\alpha \text{ and }\Psi _{n^{v}-1}\left(
x_{1}\right) +\Psi _{n^{v}}\left( x_{2}\right) \text{ for }n^{v}\beta \text{,%
}  \notag \\
B_{o,e}^{v}& \propto \Psi _{n^{v}}\left( x_{1}\right) +\Psi _{n^{v}-1}\left(
x_{2}\right) \text{ for }n^{v}\alpha \text{ and }\Psi _{n^{v}}\left(
x_{1}\right) -\Psi _{n^{v}-1}\left( x_{2}\right) \text{ for }n^{v}\beta
\text{,}  \notag \\
\text{for }x_{1}& \approx x_{2}\simeq 1/2\text{.}  \tag{10.11}
\end{align}

\begin{figure}[tbp]
\par
\begin{center}
\leavevmode
\includegraphics[width=0.8\linewidth]{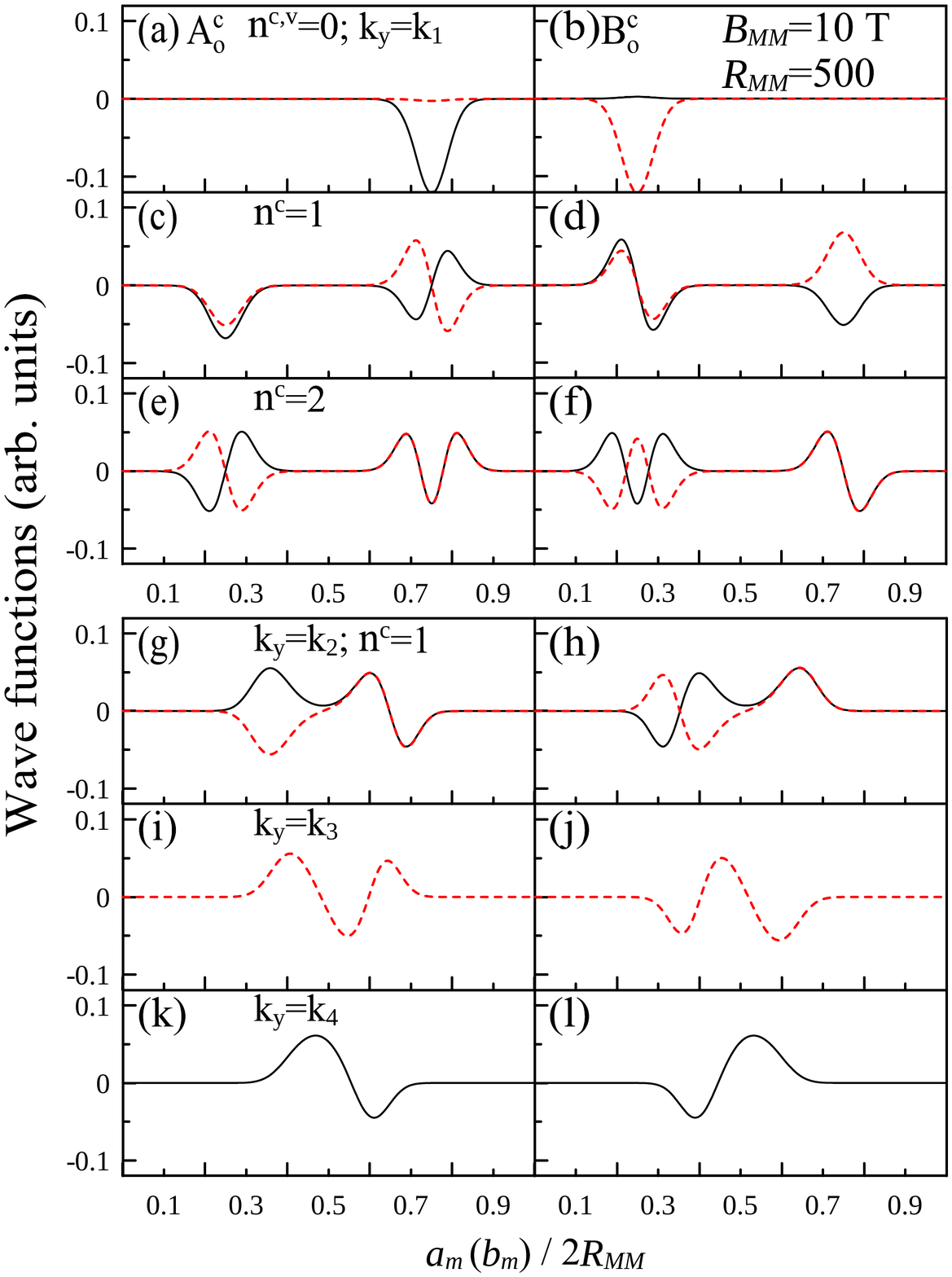}
\end{center}
\par
\textbf{Figure 10.5.} The wave functions of Quasi-Landau levels at (a)-(f) the original
band-edge state $k_{1}$ with the quantum numbers $n^{c,v}=0$, $n^{c}=1$ and $%
n^{c}=2$, (g) and (h) the split point $k_{2}$ with $n^{c}=1$, and (i)-(l)
two extra band-edge states $k_{3}$ and $k_{4}$ with $n^{c}=1$.
\end{figure}

\subsection{Optical Absorption Spectra of Quasi-Landau Levels}

Under the modulated magnetic field, the parabolic energy bands possess
several band-edge states. A wave function composed of two tight-binding
functions presents a complex overlapping behavior. The above-mentioned main
features of the electronic properties are expected to be directly reflected
in optical excitations. The low-frequency optical absorption spectra for $%
R_{MM}=500$ and $B_{MM}=10$ T, as shown in Fig. 10.6(a)\ by the black and
blue solid curves for $\widehat{\mathbf{E}}\perp \widehat{x}$ and $\widehat{%
\mathbf{E}}\parallel \widehat{x}$ respectively, exhibit rich asymmetric
peaks in the square-root divergent form. These peaks can be divided into the
principal peaks $\omega _{P}$'s and the subpeaks $\omega _{S}$'s based on
the optical excitations resulting from the original band-edge and extra
band-edge states, respectively. $\omega _{S}$'s can be further classified
into two subgroups $\omega _{S}^{a}$'s and $\omega _{S}^{b}$'s which
primarily come from the excitations of extra band-edge states $\alpha
\rightarrow \beta $ ($\beta \rightarrow \alpha $) and $\alpha \rightarrow
\alpha $ ($\beta \rightarrow \beta $), respectively. What is worth
mentioning is that the spectra for $\widehat{\mathbf{E}}\perp \widehat{x}$
and $\widehat{\mathbf{E}}\parallel \widehat{x}$ are distinct, especially for
the subpeaks $\omega _{S}$'s. The former is mainly composed of the subgroup $%
\omega _{S}^{a}$, while the latter mainly consists of the subgroup $\omega
_{S}^{b}$. This implies that the optical absorption spectra reflect the
anisotropy of the polarization direction. For the modulation along the
zigzag direction at $R_{MM}=866$ and $B_{MM}=10$ T, the absorption spectrum
(red dashed curve in Fig. 10.6(a)) shows features similar to those of the
spectrum corresponding to the armchair direction at $R_{MM}=500$ and $%
B_{MM}=10$ T. $R_{MM}=866$ for the zigzag direction and $R_{MM}=500$ for the
armchair direction possess the same period length based on the definitions $%
R_{MM}=$ $l_{MM}/3b^{\prime }$ and $R_{MM}=$ $l_{MM}/\sqrt{3}b^{\prime }$
associated with the zigzag and armchair directions, respectively. Moreover,
the anisotropic features of the modulation directions will be revealed in
the higher frequency region or the smaller modulation length.

\begin{figure}[tbp]
\par
\begin{center}
\leavevmode
\includegraphics[width=0.8\linewidth]{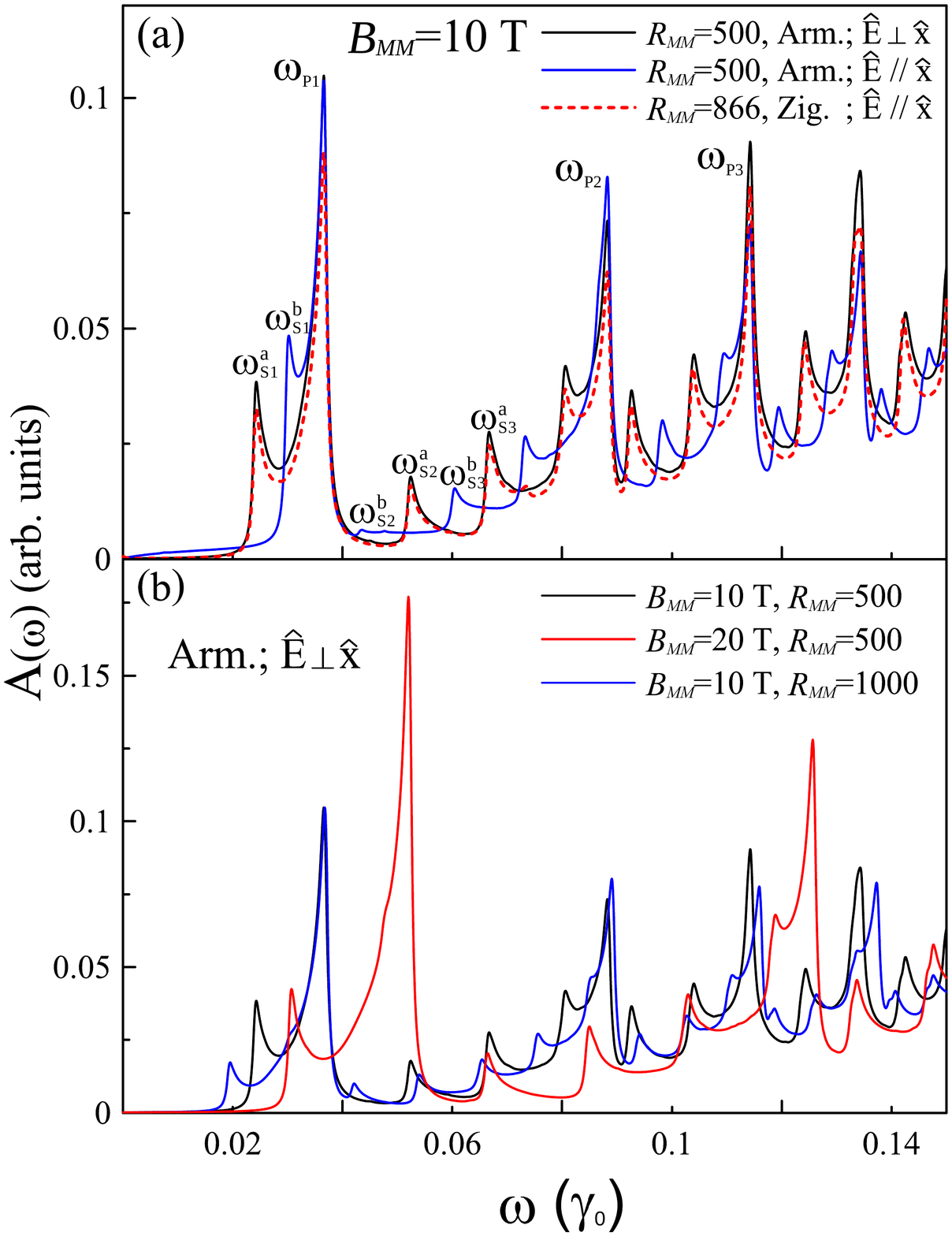}
\end{center}
\par
\textbf{Figure 10.6.} The optical absorption spectra for (a) $B_{MM}=10$ T at a fixed
periodic length with modulation and polarization along the armchair and
zigzag directions and (b) different modulation periods and field strengths
with both the modulation and polarization along the armchair direction.
\end{figure}

As the field strength rises, the peak height and frequency of the principal
peaks increase, and the peak number decreases, as shown in Fig. 10.6(b) by
the red curve for $R_{MM}=500$ and $B_{MM}=20$ T. These results mean that
the congregation of electronic states is more pronounced as the field
strength grows. In addition to the field strength, the optical-absorption
spectrum is also influenced by the modulation period. In Fig. 10.6(b), the
blue curve shows the optical spectra of $B_{MM}=10$ T for $R_{MM}=1000$. The
subpeaks strongly depend on the period, i.e., they represent different peak
heights and frequencies with the variation of $R_{MM}$. However, the
opposite is true for the principal peaks.

The peaks in the low-frequency absorption spectra can arise from the
different selection rules. Fig. 10.4 illustrates the transition channels of
the principal peaks resulting from the original band-edge states denoted as $%
\omega _{Pn}$'s\ in Fig. 10.6(a). Each $\omega _{Pn}$\ corresponds to the
transition channels from QLLs $n\rightarrow n+1$ and $n\rightarrow n+1$ at
the original band-edge state and the selection rule is represented by $%
\Delta n=|n^{c}-n^{v}|=1$ which is same as that related to LLs. The main
reason for this is that the subenvelope functions $A_{\mathbf{o}%
}^{c(v)}(x_{1})$ ($A_{\mathbf{o}}^{c(v)}(x_{2})$) and $B_{\mathbf{o}%
}^{v(c)}(x_{1})$ ($B_{\mathbf{o}}^{v(c)}(x_{2})$) associated with the
effective quantum numbers $n+1$ ($n$) and $n$ ($n+1$) have the same number
of zeros, respectively. As discussed in the former section, peaks arise in
the optical absorption spectra when the number of zeros is the same for $A_{%
\mathbf{o}}^{c(v)}$ and $B_{\mathbf{o}}^{v(c)}$ in Eq. (10.10). The subpeaks
originating from the extra band-edge states display a more complex behavior.
The excitation channels for the subpeaks $\omega _{Sn}^{a}$ and $\omega
_{Sn}^{b}$ in Fig. 10.6(a) are shown in Fig. 10.4. The subpeaks of different
selection rules, $\Delta n=0$ and $1$, come into existence simultaneously.
For example, $\omega _{S2}^{a}$ comes from the excitation channel\ $1\alpha $
$\rightarrow $ $1\beta $ ($1\beta \rightarrow 1\alpha $)\ and $\omega
_{S3}^{a}$ comes from the excitation channel $1\beta \rightarrow 2\alpha $ ($%
2\alpha \rightarrow 1\beta $). The extra selection rule $\Delta n=0$\
reflects the overlap of subenvelope functions $A_{\mathbf{o}}^{c}$($x_{1}$) (%
$B_{\mathbf{o}}^{c}$($x_{1}$)) and $A_{\mathbf{o}}^{c}$($x_{2}$) ($B_{%
\mathbf{o}}^{c}$($x_{2}$)) located around $x_{1}\approx x_{2}\approx 1/2$.
The subenvelope functions $A_{\mathbf{o}}^{c(v)}(x_{1})$ ($A_{\mathbf{o}%
}^{c(v)}(x_{2})$)\ and $B_{\mathbf{o}}^{v(c)}(x_{2})$ ($B_{\mathbf{o}%
}^{v(c)}(x_{1})$) of the effective quantum number $n$ also have the same
number of zeros at the identical position, a cause leading to the extra
selection rule $\Delta n=0$.

The frequency of principal peaks in the optical absorption spectra is worth
a closer investigation. The relation between the frequencies of the first
four principal peaks and the modulation period is shown in Fig. 10.7(a). The
$\omega _{P}$'s present a very weak dependence on the period as $R_{MM}$
becomes sufficiently\ large, whereas they exhibit a strong dependence on the
field strength. The frequencies grow with increased $B_{MM}$, as shown in
Fig. 10.7(b). The dependence of $\omega _{P}$'s on $B_{MM}$ is similar to
what is seen in the case of a uniform perpendicular magnetic field, i.e., $%
\omega _{P}$'s $\propto \sqrt{B_{MM}}$, as indicated by the red lines. The
predicted results could be verified by optical spectroscopy.\cite%
{prb25,prb24,10.2.12}

\begin{figure}[tbp]
\par
\begin{center}
\leavevmode
\includegraphics[width=0.8\linewidth]{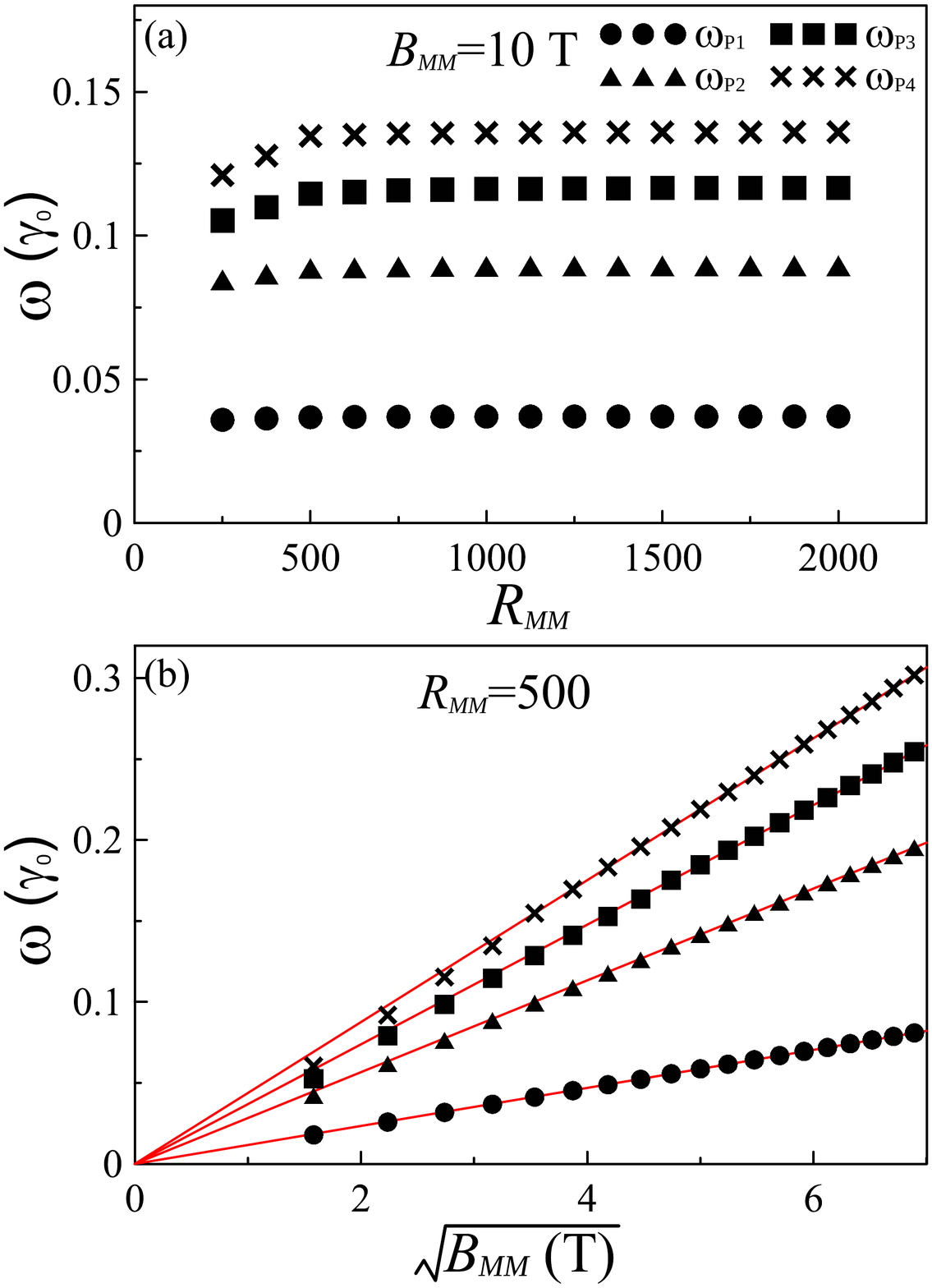}
\end{center}
\par
\textbf{Figure 10.7.} The dependence of the absorption frequency on (a) the period $%
R_{MM}$ and (b) the square root of field strength $B_{MM}$.
\end{figure}

\newpage

\section{Spatially Modulated Electric Potential}

\subsection{Oscillation Energy Subbands}

Besides the spatially modulated magnetic field, the low-energy physical
properties can also be strongly tuned by a spatially modulated electric
potential. The energy bands for $V_{ME}=0.05$ $\gamma _{0}$ and $R_{ME}=500$
are shown in Fig. 10.8. The unoccupied conduction subbands are symmetric to
the occupied valence subbands about $E_{F}$. The parabolic subbands are
nondegenerate and oscillate near $k_{y}=2/3$. There exists on intersection
where two parabolic subbands cross each other at $E_{F}$. Each subband has
several band-edge states, which lead to the prominent peaks in the DOS and
optical absorption spectra. For convenience, these band-edge states are
further divided into two categories called $\mu $ and $\nu $ states, as
indicated in Fig. 10.8. The two $\mu $ ($\nu $) states at the left- and
right-hand sites of $k_{y}=2/3$ might have a small difference in energies;
that is, parabolic bands might be bilaterally asymmetric about $k_{y}=2/3$.
Not far away from $k_{y}=2/3$, the energy subbands with linear dispersions
intersect at $E_{F}$, preserving more Fermi-momentum states and forming
several Dirac cones. Moreover, the number of Fermi-momentum states or Dirac
cones increases with the potential strength and modulation period.

\begin{figure}[tbp]
\par
\begin{center}
\leavevmode
\includegraphics[width=0.8\linewidth]{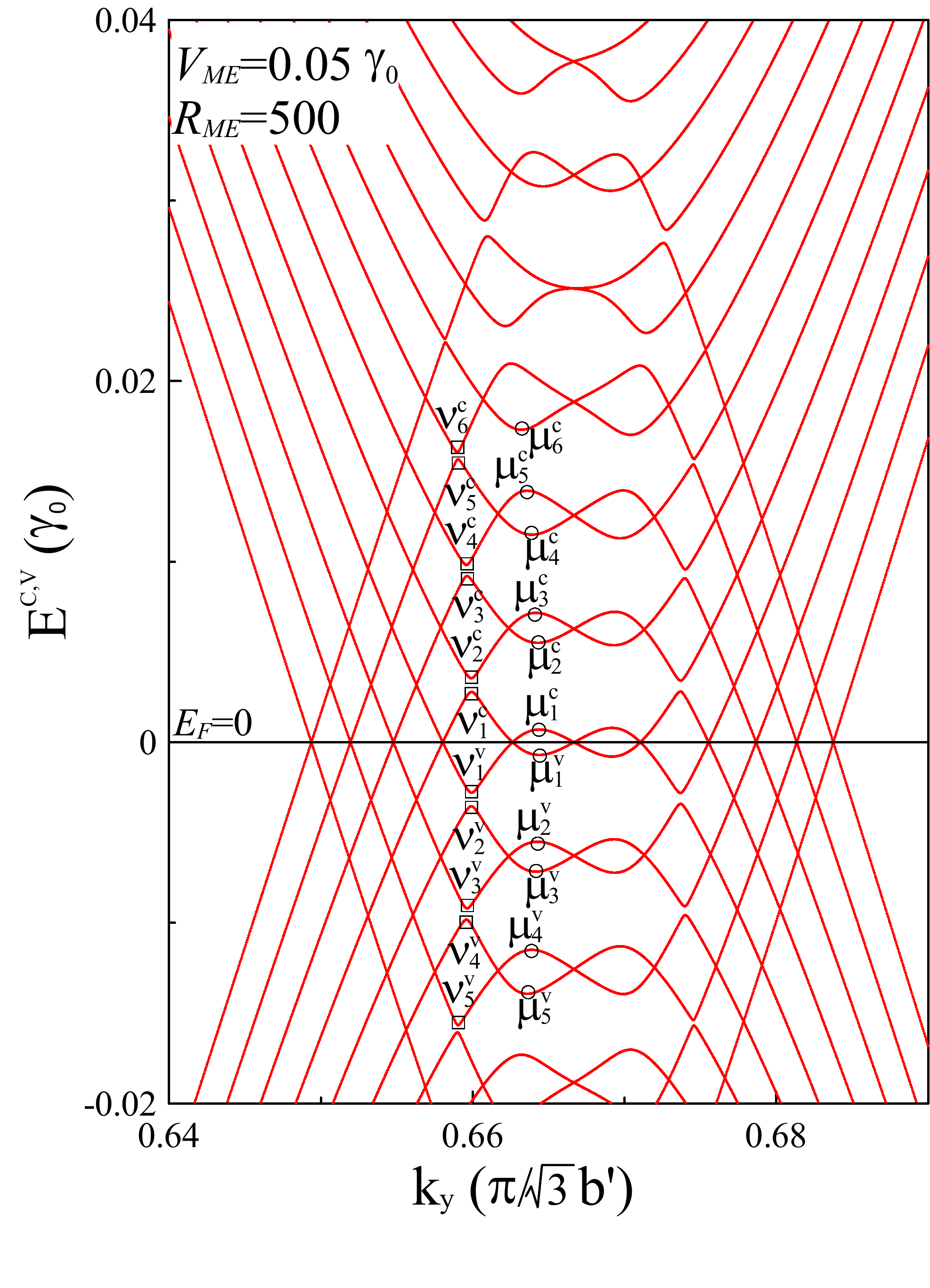}
\end{center}
\par
\textbf{Figure 10.8.} The energy dispersions for the modulated electric potential along
the armchair direction with $R_{MM}=500$ and $V_{ME}=0.05\gamma _{0}$.
\end{figure}

The optical absorption spectrum for $R_{ME}=500$ and $V_{ME}=0.05$ $\gamma
_{0}$ along the armchair direction, as shown in Fig. 10.9 by the black solid
curve, exhibits two groups of prominent peaks, $\Sigma _{n}$'s and $\Upsilon
_{n}$'s. They are mainly due to the optical excitations from $\mu _{n}^{v}$
to $\mu _{n+1}^{c}$ ($\mu _{n+1}^{v}$ to $\mu _{n}^{c}$) and $\mu _{n}^{v}$
to $\mu _{n+2}^{c}$ ($\mu _{n+2}^{v}$ to $\mu _{n}^{c}$), respectively.
Moreover, with regard to the peak intensity, the peaks $\Sigma _{n}$'s ($%
\Upsilon _{n}$'s) can be further divided into two subgroups. For example,
the peak heights of $\Sigma _{1}$, $\Sigma _{3}$; $\Sigma _{5}$,
respectively, resulting from the transitions of $\mu _{1}^{v}$ to $\mu
_{2}^{c}$ ($\mu _{2}^{v}$ to $\mu _{1}^{c}$), $\mu _{3}^{v}$ to $\mu
_{4}^{c} $ ($\mu _{4}^{v}$ to $\mu _{3}^{c}$); $\mu _{5}^{v}$ to $\mu
_{6}^{c}$ ($\mu _{6}^{v}$ to $\mu _{5}^{c}$) are very low, while the peaks $%
\Sigma _{2}$, $\Sigma _{4}$; $\Sigma _{6}$ originating from the excitations $%
\mu _{2}^{v}$ to $\mu _{3}^{c}$ ($\mu _{3}^{v}$ to $\mu _{2}^{c}$), $\mu
_{4}^{v}$ to $\mu _{5}^{c}$ ($\mu _{5}^{v}$ to $\mu _{4}^{c}$); $\mu
_{6}^{v} $ to $\mu _{7}^{c} $ ($\mu _{7}^{v}$ to $\mu _{6}^{c}$) present
much stronger intensities than the peaks $\Sigma _{1}$, $\Sigma _{3}$ and $%
\Sigma _{5}$. That is to say, the peak of $\Sigma _{2n}$'s are higher than
those of $\Sigma _{2n-1}$'s in the group $\Sigma _{n}$. The peaks of $%
\Upsilon _{n}$'s exhibit similar features to those of $\Sigma _{n}$'s. For
instance, peaks $\Upsilon _{1}$, $\Upsilon _{3}$; $\Upsilon _{5}$,
respectively arising from the transitions of $\mu _{1}^{v}$ to $\mu _{3}^{c}$
($\mu _{3}^{v}$ to $\mu _{1}^{c}$), $\mu _{3}^{v}$ to $\mu _{5}^{c}$ ($\mu
_{5}^{v}$ to $\mu _{3}^{c} $); $\mu _{5}^{v} $ to $\mu _{7}^{c}$ ($\mu
_{7}^{v}$ to $\mu _{5}^{c}$), own the peaks with very weak intensities. In
contrast to $\Upsilon _{1}$, $\Upsilon _{3}$; $\Upsilon _{5}$, the peak
intensities of $\Upsilon _{2}$ and $\Upsilon _{4}$ resulting from the
excitations $\mu _{2}^{v}$ to $\mu _{4}^{c}$ ($\mu _{4}^{v}$ to $\mu
_{2}^{c} $) and $\mu _{4}^{v}$ to $\mu _{6}^{c}$ ($\mu _{6}^{v}$ to $\mu
_{4}^{c}$) are relatively stronger. Furthermore, the $\mu $ and $\nu $
states lead to different contributions to the two kinds of optical
absorption peaks. Most peaks originating from the two different band-edge
states have nearly the same frequencies, while the peak intensities are not
the same. The blue and red curves correspond to the optical absorption
spectra which contains only the excitations of $\mu $ and $\nu $ states,
respectively. Except for the peak $\Sigma _{2}$ with comparable
contributions which are attributed to the transitions of $\mu $ and $\nu $
states, the other peaks with different contributions from the two states
have nearly the same frequency. The peaks from the $\mu $ states exhibit
much stronger intensities than those from the $\nu $ states. In other words,
peaks in the optical absorption spectrum mainly result from excitations of
the $\mu $ states.

\begin{figure}[tbp]
\par
\begin{center}
\leavevmode
\includegraphics[width=0.8\linewidth]{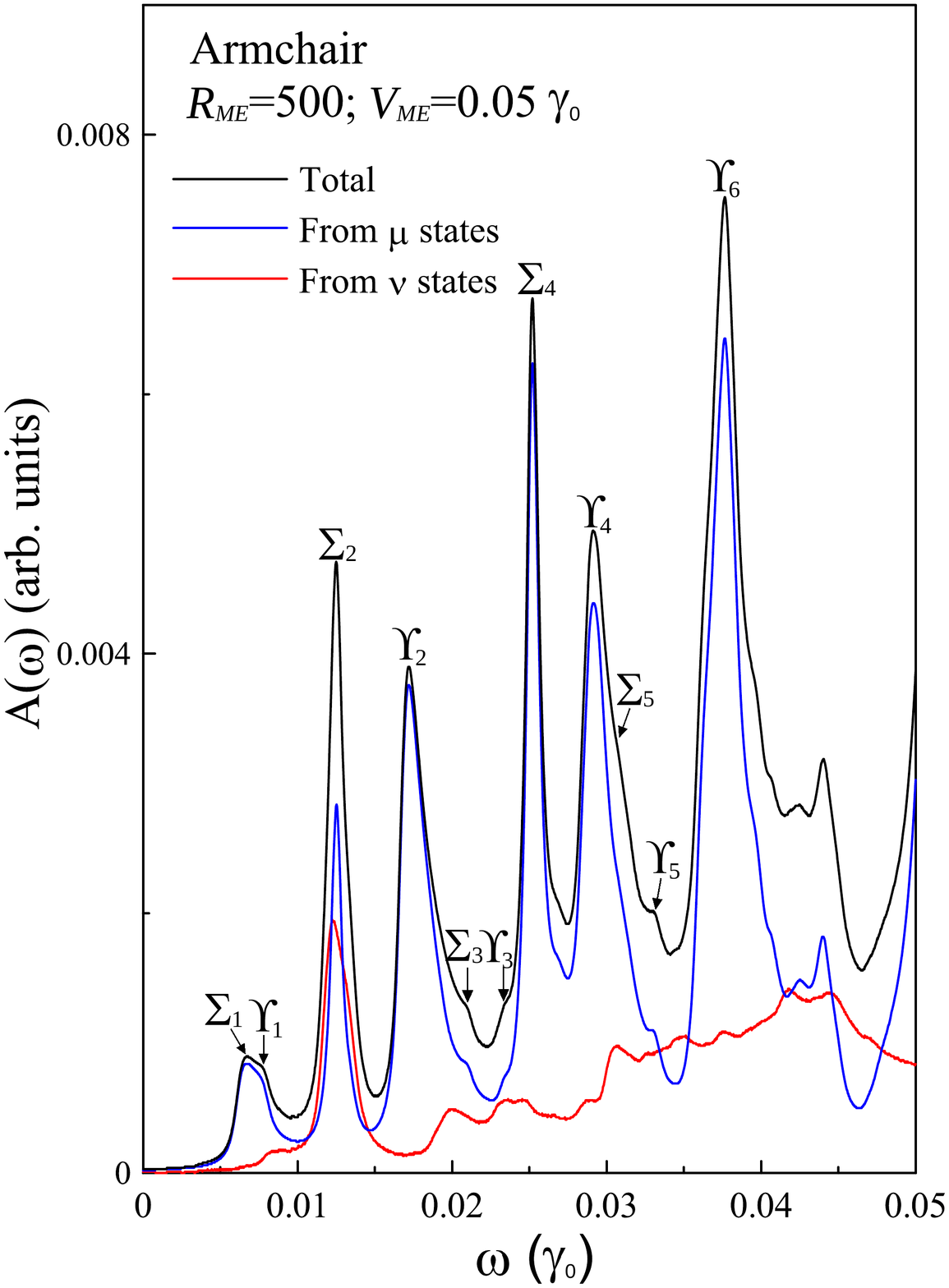}
\end{center}
\par
\textbf{Figure 10.9.} The optical absorption spectra corresponding to Fig. 10.8, which
includes the contributions from the $\mu $ and $\nu $ states, respectively.
\end{figure}

\subsection{Anisotropic Optical Absorption Spectra}

The polarization direction and the strength, period and direction of the
modulating electric field strongly affect the features of the optical
absorption spectrum. The spectra associated with $\widehat{\mathbf{E}}\perp
\widehat{x}$ (black solid curve) and $\widehat{\mathbf{E}}\parallel \widehat{%
x}$ (red solid curve) for $R_{ME}=500$ and $V_{ME}=0.05$ $\gamma _{0}$ along
the armchair direction and $R_{ME}=866$ and $V_{ME}=0.05$ $\gamma _{0}$
along the zigzag direction (blue solid curve) are shown in Fig. 10.10(a) for
a comparison. Compared with the results of $\widehat{\mathbf{E}}\perp
\widehat{x}$ and $\widehat{\mathbf{E}}\parallel \widehat{x}$, the peak
structures related to the two polarization directions are totally different,
which reflect the anisotropic behavior of the polarization direction.
Similarly, the anisotropy of the modulation directions is reflected by that
the absorption spectra corresponding to the armchair and zigzag directions
display distinct features, i.e., the anisotropic behavior of the
polarization directions are more obvious than that in the MM case. With
increasing the modulation strength to $V_{ME}=0.1\gamma _{0}$ (red solid
curve in Fig. 10.10(b)), the results show that the peak intensity strongly
depends on $V_{ME}$, but their relationship is not straight forward. For the
modulation period, the spectra at a larger $R_{ME}=1000$ (blue solid curve)
along the armchair direction present features diverse to those in the
spectra at $R_{ME}=1000$. The peak number grows and the peak intensities
decay with a increase of the period. A redshift occurs in longer periods.
For example, the peak frequencies $\Sigma _{1}$, $\Sigma _{3}$; $\Sigma _{5}$%
, as indicated in black and green curves, are\ almost reduced to half of the
original ones when the modulation period is enlarged from 500 to 1000.

\begin{figure}[tbp]
\par
\begin{center}
\leavevmode
\includegraphics[width=0.8\linewidth]{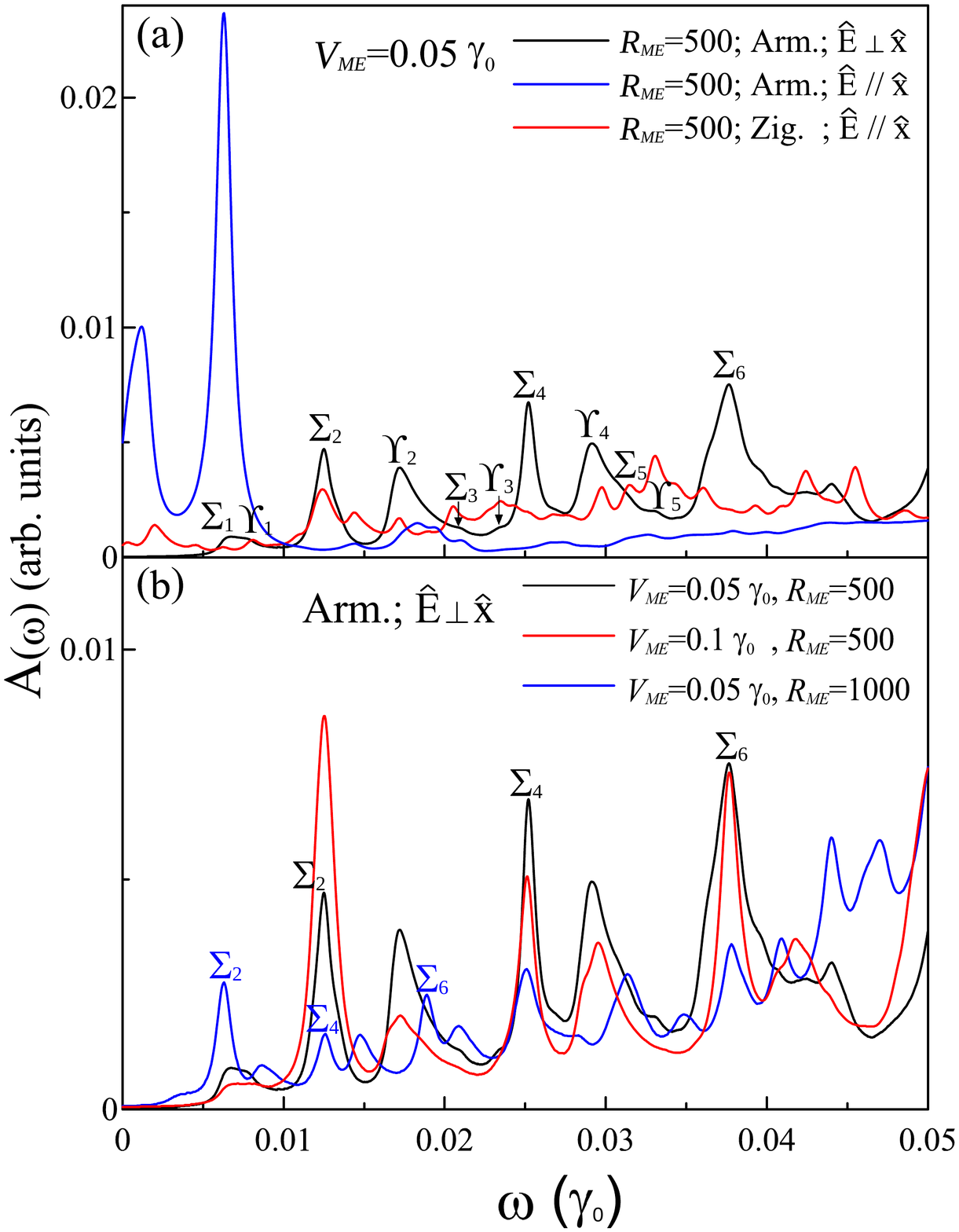}
\end{center}
\par
\textbf{Figure 10.10.} The optical absorption for (a) $V_{ME}=0.05\gamma _{0}$ at a
fixed periodic length with modulation and polarization along the armchair
and zigzag directions and (b) different modulation periods and field
strengths with both the modulation and polarization along the armchair
direction.
\end{figure}

The optical absorption spectra in the ME case do not reveal certain
selection rules. This is due to the fact that the amplitudes $A_{\mathbf{o}%
}^{c,v}$ and $B_{\mathbf{o}}^{v,c}$ of the wave functions do not exist a
simple relationship similar to that in the UM and MM cases. The wave
functions in the modulated electric potential are no longer distributed
around the center location; rather, they display standing-wave-like features
in the primitive unit cell and are distributed over the entire primitive
cell, as shown in Fig. 10.11. However, the wave functions of the edge-states
$\mu $ and $\nu $ exhibit irregular behavior such as disordered\ numbers of
zero points, asymmetric spatial distributions, and random oscillations.
These irregular waveforms might result from different site energies for the
carbon atoms in the modulated electric potential.

\begin{figure}[tbp]
\par
\begin{center}
\leavevmode
\includegraphics[width=0.8\linewidth]{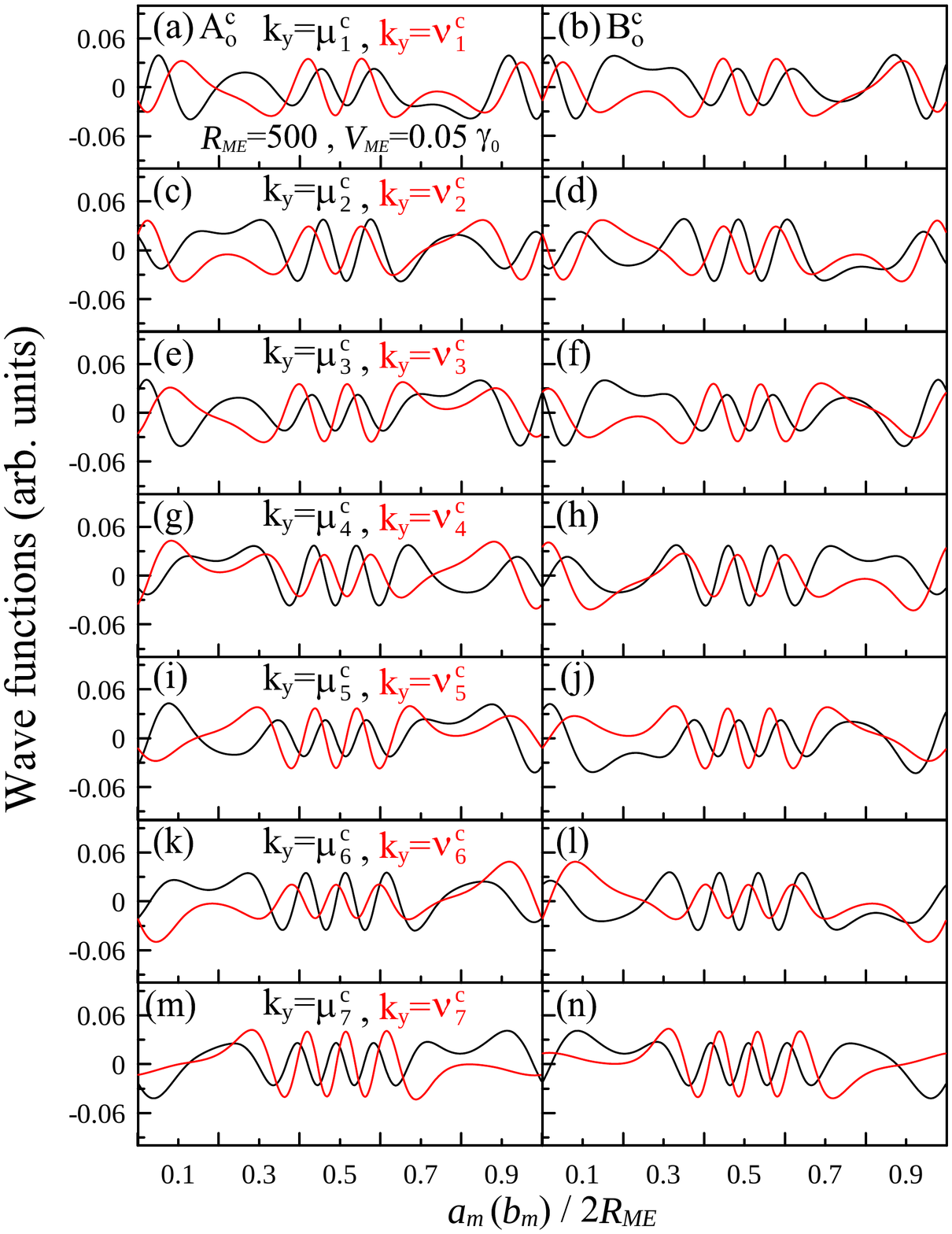}
\end{center}
\par
\textbf{Figure 10.11.} The wave functions at different band-edge states, the $\mu
_{i}^{c}$ and $\nu _{i}^{c}$ states for $i=1\symbol{126}7$.
\end{figure}

\newpage

\section{Uniform Magnetic Field Combined with Modulated Magnetic Field}

\subsection{Landau Level Spectra Broken by Modulated Magnetic Fields}

A further discussion of graphene in a composite field, the UM-MM case, is
presented in this section. The main characteristics of the LLs at $B_{UM}=5$
T are affected by the modulated magnetic field ($B_{MM}=1$ T and $R_{MM}=395$%
), as shown in Fig. 10.12(a) by the black curves. The LL with $n^{c,v}=0$ at
$E_{F}=0$ remains the same features of the UM case. On the other hand, each
dispersionless LL with $n^{c,v}\geqslant 1$ splits into two parabolic
subbands with double degeneracy. The subbands possess two kinds of band-edge
states, $n^{c,v}\zeta $ and $n^{c,v}\eta $, which correspond to the minimum
field strength $B_{UM}-B_{MM}$ and maximum field strength $B_{UM}+B_{MM}$,
respectively. The surrounding electronic states at $n^{c,v}\eta $ congregate
more easily, which results in the smaller band curvature. Comparably fewer
states congregate at $n^{c,v}\zeta $, and the resulting band curvature is
larger. Increasing $B_{MM}$ induces more complex energy spectra, as shown in
Fig. 10.12(b) for $R_{MM}=395$ and $B_{MM}=5$ T. The parabolic subbands with
$n^{c,v}\geqslant 1$ display wider oscillation amplitudes, stronger energy
dispersions, and greater band curvatures. The largest and smallest band
curvatures occur at the local minima $n^{c,v}\zeta $ and local maxima $%
n^{c,v}\eta $ states, respectively. The subband amplitudes are nearly
linearly magnified by $B_{MM}$ as $B_{MM}\leq B_{UM}$. It is noticeable that
neither the minima of the conduction bands nor the maxima of the valence
bands exceed $E_{F}=0$ even for $B_{MM}$ much larger than $B_{UM}$, as shown
in Fig. 10.12(c) for $R_{MM}=395$ and $B_{MM}=40$ T. Thus no overlap exists
between the conduction and valence bands, regardless of the modulation
strength. With further increasing modulated field strength as $B_{MM}\gg
B_{UM}$, the electronic structures are expected to approach to those in the
MM case.

\begin{figure}[tbp]
\par
\begin{center}
\leavevmode
\includegraphics[width=0.8\linewidth]{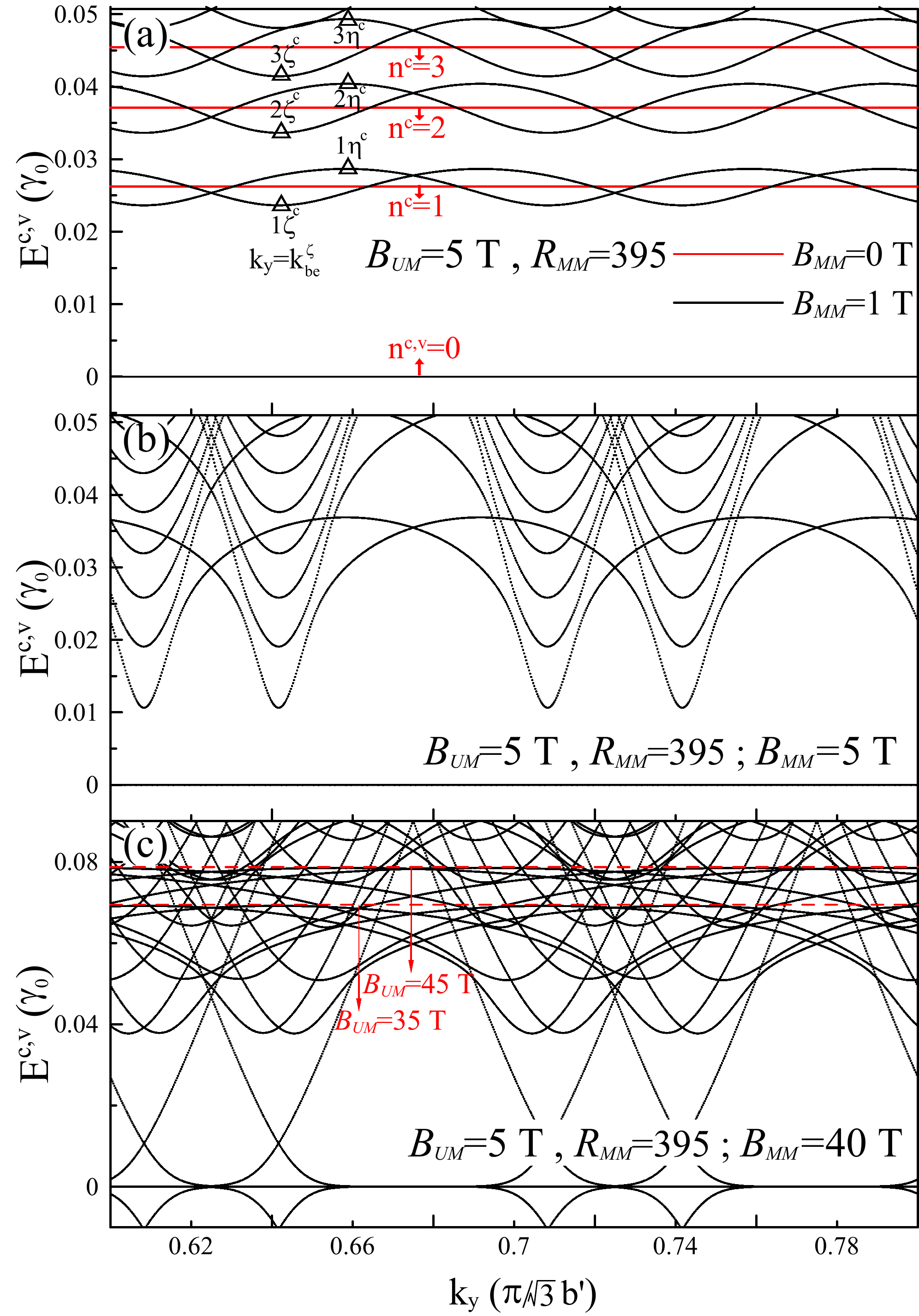}
\end{center}
\par
\textbf{Figure 10.12.} The energy dispersions for (a) the uniform magnetic field $%
B_{UM}=5$ T by the red curves and the composite field $B_{UM}=5$ T combined
with $R_{MM}=395$ and $B_{MM}=1$ T by the black curves, (b) $B_{UM}=5$ T
combined with $R_{MM}=395$ and $B_{MM}=5$ T, and (c) $B_{UM}=5$ T combined
with $R_{MM}=395$ and $B_{MM}=40$ T. All modulated fields are applied along
the armchair direction.
\end{figure}

\subsection{Symmetry Broken of Landau Level Wave Functions}

The LL wave functions modified by the modulated magnetic field are shown in
Fig. 10.13. The spatial distributions corresponding to $k_{be}^{\zeta }$,
labeled in Fig. 10.12, exhibit slightly broadened and reduced amplitudes, as
indicated by the black curves in Figs. 10.13(a)-(d) for $B_{MM}=1$ T.
However, the spatial symmetry and the location centers of the wave functions
remain unchanged. Under the influence of a small $B_{MM}$, the simple
relation between $A_{\mathbf{o}}^{c,v}$ and $B_{\mathbf{o}}^{c,v}$ of the
wave functions is almost preserved. However, a stronger modulation strength
results in greater spatial changes of the wave functions, as shown in Figs.
10.13(e)-(f) for $B_{UM}=B_{MM}=5$ T. The increased broadening and asymmetry
of the spatial distributions of the wave functions at $n^{c,v}=0$ are
revealed. However, the spatial distributions with $n^{c,v}\geqslant 1$ are
only widened (i.e., $n^{c}=1$ in Figs. 10.12(g) and (h)), but the spatial
symmetry is retained. With increasing $B_{MM}$, as shown in Figs
10.13(i)-(l) for $B_{MM}=40$ T, the symmetry of the wave functions with $%
n^{c,v}=0$ is recovered and one can expect that the main features of the
wave functions will become similar to those in the MM case. Obviously, the
electronic properties show critical changes as $B_{MM}$ equals $B_{UM}$,
which should be reflected to the optical properties.

\begin{figure}[tbp]
\par
\begin{center}
\leavevmode
\includegraphics[width=0.8\linewidth]{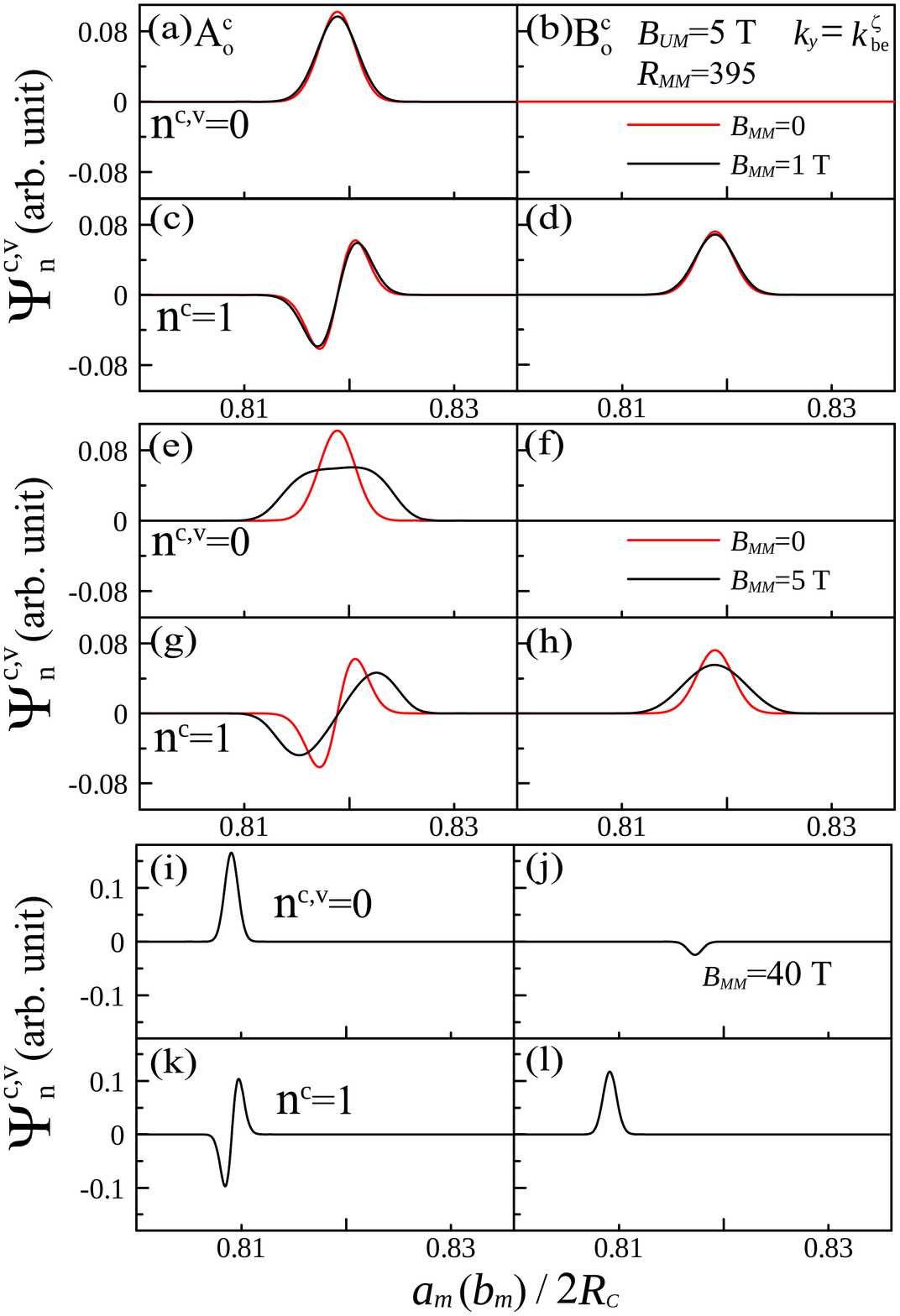}
\end{center}
\par
\textbf{Figure 10.13.} The wave functions with $n^{c,v}=0$ and $n^{c}=1$ at $%
k_{be}^{\zeta }$ for (a)-(d) the uniform magnetic field $B_{UM}=5$ T by the
red curves and the composite field $B_{UM}=5$ T combined with $R_{MM}=395$
and $B_{MM}=1$ T by the black curves, (e)-(h) $B_{UM}=5$ T by the red curves
and $B_{UM}=5$ T combined with $R_{MM}=395$ and $B_{MM}=5$ T by the black
curves, and (i)-(l) $B_{UM}=5$ T combined with $R_{MM}=395$ and $B_{MM}=40$
T.
\end{figure}

\subsection{Magneto-Optical Absorption Spectra with Extra Selection Rules}

The optical absorption spectra corresponding to Fig. 10.13 are shown in
Figs. 10.14 and 10.15. In Fig. 10.14(a), the absorption spectra
corresponding to the UM-MM case at $B_{UM}=5$ T with $R_{MM}=395$ and $%
B_{MM}=1$ T and the UM case at $B_{UM}=5$ T are shown together for a
comparison. The red curves coming from the LLs at $B_{UM}=5$ T display
delta-function-like peaks $\omega _{LL}^{nn^{\prime }}$\ with the selection
rule $\Delta n=1$. However, the modulated magnetic field modifies each
delta-function-like peak into two split square-root-divergent peaks, $\omega
_{\zeta }^{nn^{\prime }}$ and $\omega _{\eta }^{nn^{\prime }}$, as shown by
the black curves. Each $\omega _{\zeta }^{nn^{\prime }}$ ($\omega _{\eta
}^{nn^{\prime }}$) originates from the transitions of $n\zeta \rightarrow
n+1\zeta $ and $n+1\zeta \rightarrow n\zeta $ ($n\eta \rightarrow n+1\eta $
and $n+1\eta \rightarrow n\eta $) and its absorption frequency is same as
that generated from the LLs at $B_{UM}-B_{MM}=4$ T ($B_{UM}-B_{MM}=6$ T).
These absorption peaks obey a selection rule, $\Delta n=1$,\ similar to that
in the UM case.

\begin{figure}[tbp]
\par
\begin{center}
\leavevmode
\includegraphics[width=0.8\linewidth]{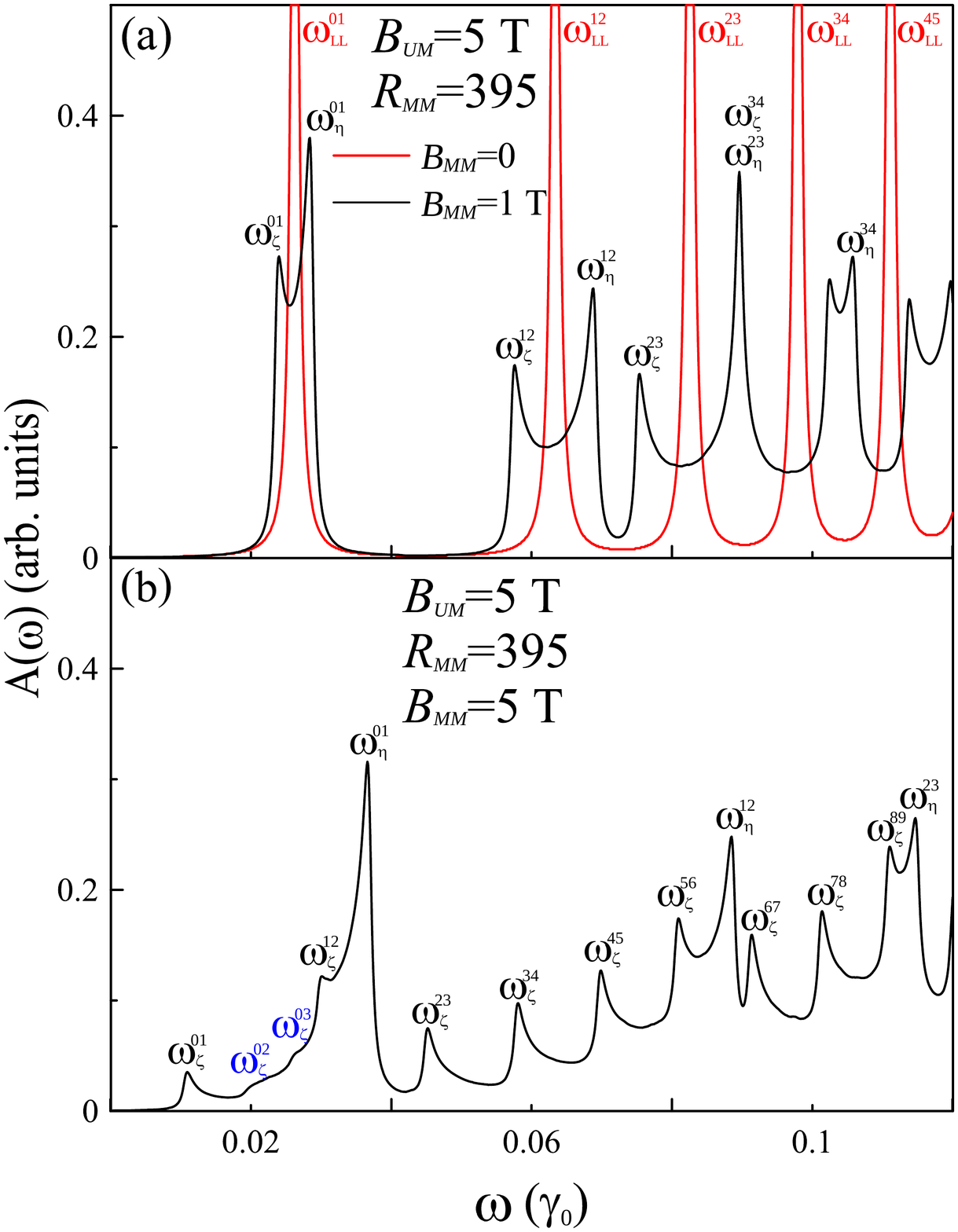}
\end{center}
\par
\textbf{Figure 10.14.} The optical absorption spectra corresponding to (a) the uniform
magnetic field $B_{UM}=5$ T by the red curve and the composite field $%
B_{UM}=5$ T combined with $R_{MM}=395$ and $B_{MM}=1$ T by the black curve
and (b) the composite field $B_{UM}=5$ T combined with $R_{MM}=395$ and $%
B_{MM}=5$ T.
\end{figure}

With increasing the modulated field strength as $B_{MM}=B_{UM}=5$ T, the
absorption spectrum has evident variety, as shown in Fig. 10.14(b). In
addition to the peaks $\omega _{\zeta }^{nn^{\prime }}$ and $\omega _{\eta
}^{nn^{\prime }}$ with the selection rule $\Delta n=1$, two extra peaks with
$\Delta n=2$ and $3$, $\omega _{\zeta }^{02}$ and $\omega _{\zeta }^{03}$,
are generated. These two peaks do reflect the fact that the wave functions
of the LLs with $n^{c,v}=0$ are destroyed by the modulated magnetic field.
As the modulated field strength further raises to $B_{MM}=40$ T (red dashed
curves Fig. 10.15), the spectrum displays some features similar to those of
the spectrum in the MM case at $R_{MM}=395$ and $B_{MM}=40$ T (black solid
curves in Fig. 10.15), i.e., the principal peaks $\omega _{P}$'s and the
subpeaks $\omega _{S}$'s in the MM case are also shown in the UM-MM case as $%
B_{MM}>B_{UM}$. Moreover, the subpeaks features, which are associated with
the positions at the net field strength equal to zero, are almost the same
in both the MM and UM-MM case. The principal peaks, however, possess a pair
structure with $\omega _{Pn}^{-}$ and $\omega _{Pn}^{+}$, which respectively
correspond to two different field strengths, $\left\vert
B_{UM}-B_{MM}\right\vert =35$ T and $\left\vert B_{UM}+B_{MM}\right\vert =45$
T, and thus the difference between two field strengths lead to distinct
absorption frequencies. For $B_{MM}\gg B_{UM}$, one can anticipate that the
frequency discrepancy between the pair $\omega _{Pn}^{-}$ and $\omega
_{Pn}^{+}$ becomes very small and then they will merge into one single peak,
$\omega _{Pn}$, i.e., the absorption spectrum restores to that in the pure
MM case.

\begin{figure}[tbp]
\par
\begin{center}
\leavevmode
\includegraphics[width=0.8\linewidth]{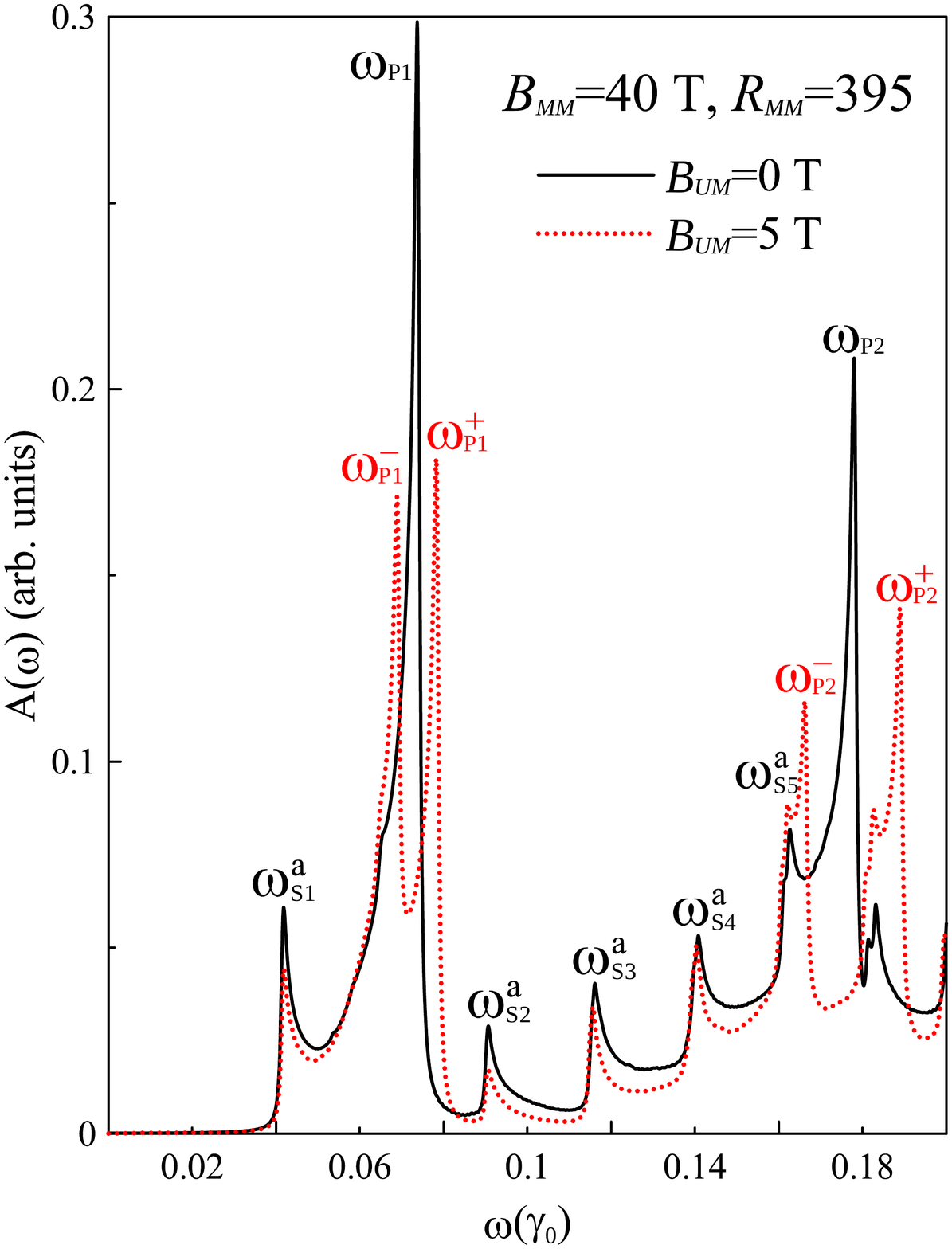}
\end{center}
\par
\textbf{Figure 10.15.} The optical absorption spectra corresponding to the composite
field $B_{UM}=5$ T combined with $R_{MM}=395$ and $B_{MM}=40$ T by the red
dotted curve and the pure modulated magnetic field $R_{MM}=395$ and $%
B_{MM}=40$ T by the black curve.
\end{figure}

The dependence of the absorption frequency on the modulated field strength
is shown in Fig. 10.16 for $B_{MM}\leq 5$ T. In the range of $B_{MM}\leq
B_{UM}$, each of absorption peaks $\omega _{\zeta }^{nn^{\prime }}$ and $%
\omega _{\eta }^{nn^{\prime }}$ is linearly dependent on $B_{MM}$. This
reflects the fact that the subband amplitudes are nearly linearly magnified
by $B_{MM}$ within the range. However, in the higher absorption frequency
region or the field range of $B_{MM}>B_{UM}$, the linear-dependence
relationship will be broken since the subbands become overlapping and the
subband amplitudes are not linearly magnified by $B_{MM}$ anymore.

\begin{figure}[tbp]
\par
\begin{center}
\leavevmode
\includegraphics[width=0.8\linewidth]{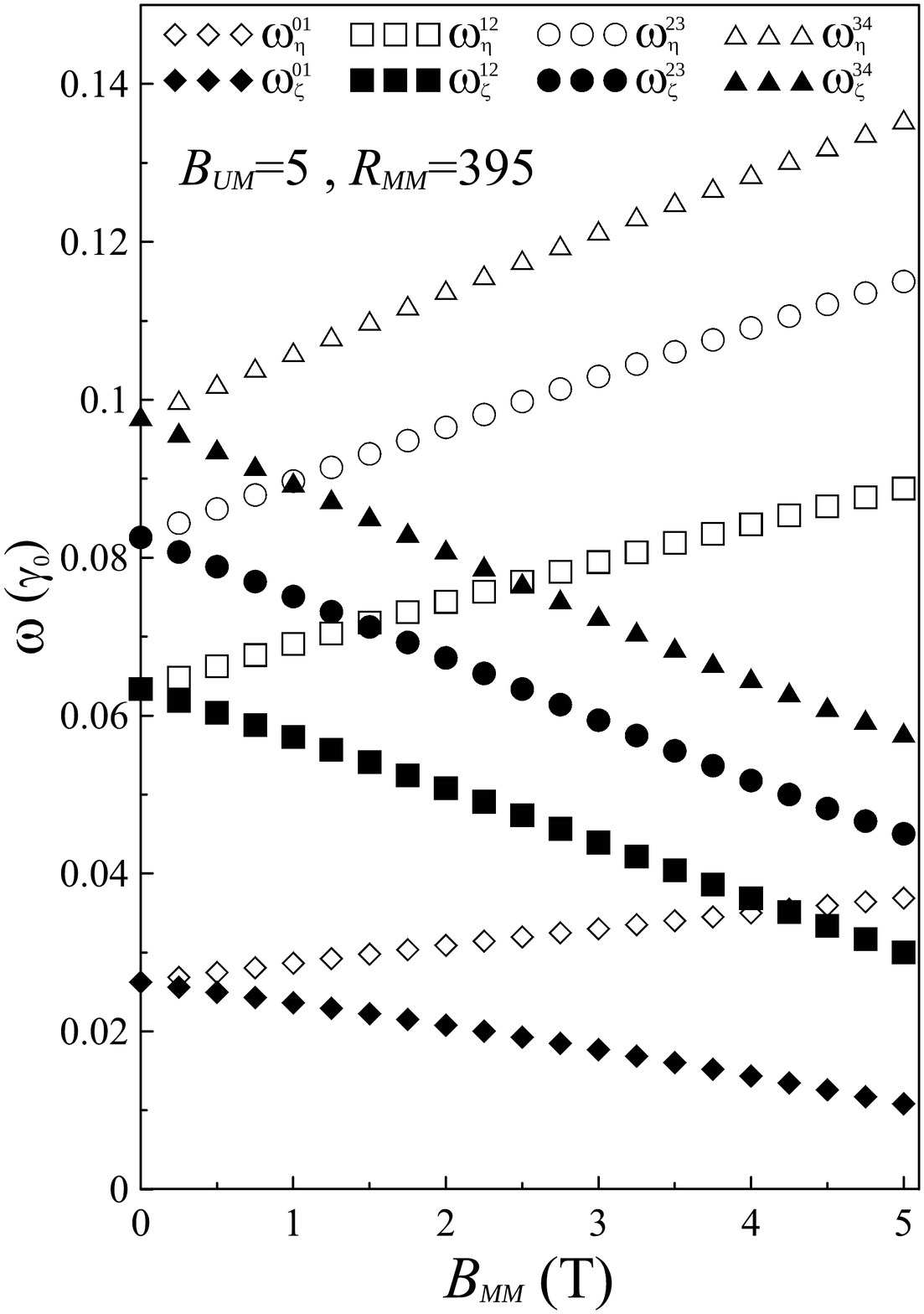}
\end{center}
\par
\textbf{Figure 10.16.} The dependence of absorption frequencies, $\omega _{\eta
}^{nn^{\prime }}$ and $\omega _{\zeta }^{nn^{\prime }}$ with $\left\vert
\Delta n\right\vert =\left\vert n-n^{\prime }\right\vert =1$, on the
modulated strength $B_{MM}$.
\end{figure}

\newpage

\section{Uniform Magnetic Field Combined with Modulated Electric Potential}

\subsection{Landau Level Spectra Broken by Modulated Electric Potentials}

Compared with the situation in a modulated magnetic field, a modulated
electric potential creates distinct effects on the LLs, as shown in Fig.
10.17(a) by the black curves for . The 0D LLs become the 1D sinusoidal
energy subbands when electronic states are affected by the periodic electric
potential. Each LL with four-fold degeneracy is split into two doubly
degenerate Landau subbands (LSs), as shown in Fig. 10.17(a) by the black
curves. Each LS owns two types of extra band-edge states, $k_{be}^{\varkappa
}{}$ and $k_{be}^{\varrho }$. For the conduction (valence) LSs, the
band-edge states of $k_{be}^{\varkappa }{}$ and $k_{be}^{\varrho }$ ($%
k_{be}^{\varrho }$ and $k_{be}^{\varkappa }{}$) are, respectively, related
to the wave functions, which possess a localization center at the minimum
and maximum electric potentials (discussed in the ME case). It should be
noted that the two LSs of $n^{c,v}=0$ only have the $\varrho $-type
band-edge states. Under a small modulation strength, the energy spacings ($%
E_{s}$'s) or band curvatures for both $k_{be}^{\varkappa }{}$ and $%
k_{be}^{\varrho }$\ are almost the same, where $E_{s}$ is the spacing
between a LS and a LL at $k_{be}^{\varkappa }$ or $k_{be}^{\varrho }$. On
the other hand, when the modulation strength is sufficiently large (e.g., $%
V_{ME}=0.02$ $\gamma _{0}$ in Fig. 10.17(b)), the energy dispersions of LSs
are relatively strong and $E_{s}$ decreases with increasing state energies,
i.e., the oscillations of LSs decline with an increase of $n^{c,v}$. In
comparison to the $k_{be}^{\varrho }$ state, the $k_{be}^{\varkappa }{}$
state owns the smaller energy spacing and band curvature. Such differences
are associated with the localization of the wave function within the
potential well.

\begin{figure}[tbp]
\par
\begin{center}
\leavevmode
\includegraphics[width=0.8\linewidth]{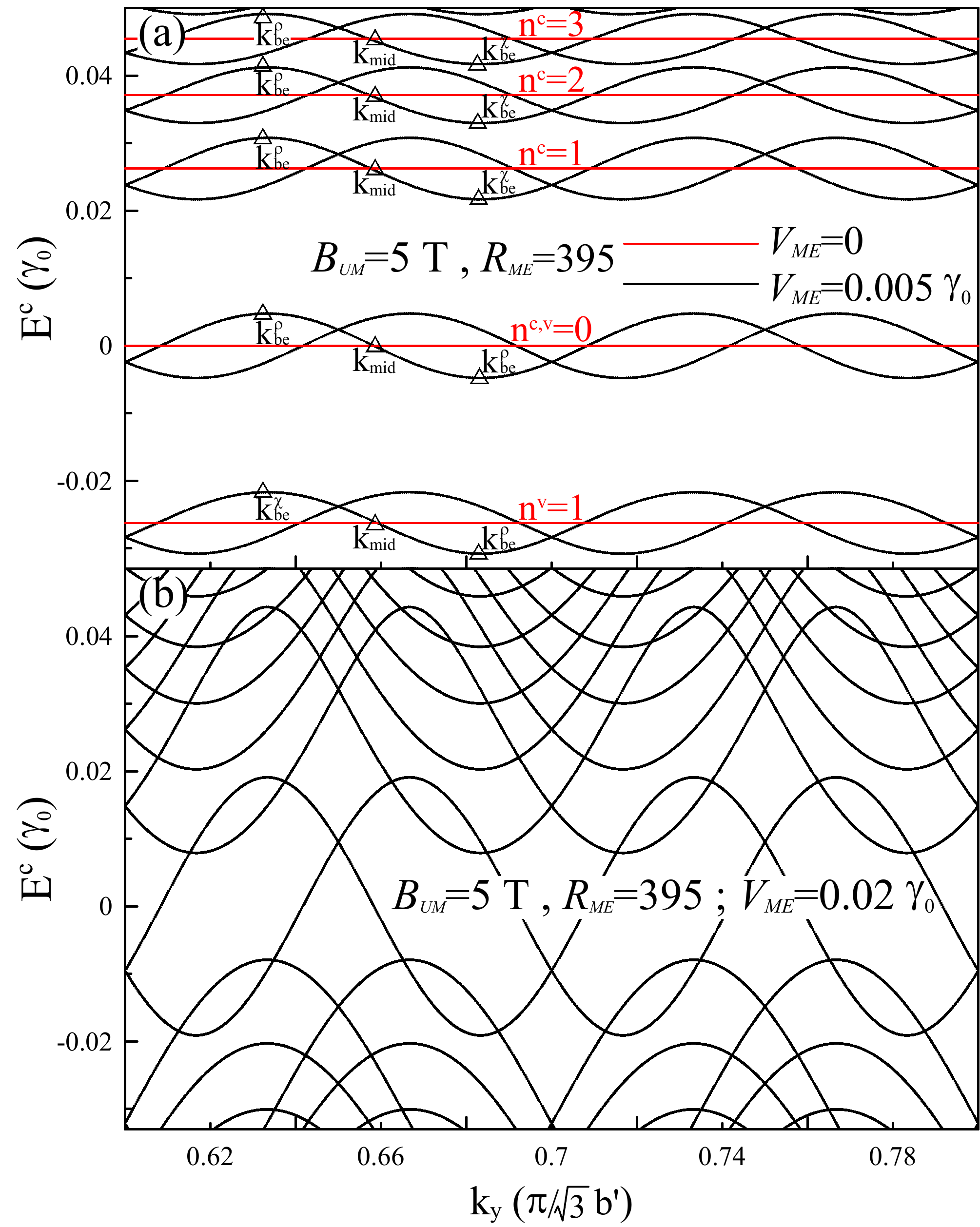}
\end{center}
\par
\textbf{Figure 10.17.} The energy dispersions for (a) the uniform magnetic field $%
B_{UM}=5$ T by the red curves and the composite field $B_{UM}=5$ T combined
with $R_{MM}=395$ and $V_{ME}=0.005\gamma _{0}$ by the black curves, (b) $%
B_{UM}=5$ T combined with $R_{MM}=395$ and $V_{ME}=0.02\gamma _{0}$. All
modulated fields are applied along the armchair direction.
\end{figure}

\subsection{Landau Level Wave Functions Broken by Modulated Electric
Potentials}

The main features of the LL wave functions are altered by the modulated
electric potential. The illustrated wave functions at the band-edge states
and the midpoint ($k_{mid}$, indicated in Fig. 10.17(a)) between two
band-edge states are used to examine the modulation effects. The wave
functions at $k_{mid}$\ are modified by $V_{ME}(x)$, as shown in Figs.
10.18(a)-(h) for $B_{UM}=5$ T and $R_{ME}=395$ at $V_{ME}=0$, $0.005$ and $%
0.02$ $\gamma _{0}$. $A_{\mathbf{o}}^{c,v}$ of $n^{c,v}=0$ is slightly
reduced, while $B_{\mathbf{o}}^{c,v}$ of $n^{c,v}=0$ is slightly increased
(Figs. 10.18(a) and 10.18(b)) after $V_{ME}$ is introduced. This means that
carriers are transferred between the $a$- and $b$-sublattices. With an
increasing $n^{c,v}$, the spatial distribution symmetry of the LL wave
functions is broken. The conduction and valence wave functions are,
respectively, shifted toward the $+\widehat{x}$ and\ $-\widehat{x}$\
directions, as shown in Figs. 10.18(e)-(h) for example. The proportionality
relationship between $A_{\mathbf{o}}^{c,v}$ of $n^{c,v}$ and $B_{\mathbf{o}%
}^{c,v}$ of $n^{c,v}+1$ no longer exists, and neither do the relationships $%
A_{\mathbf{o}}^{c}=A_{\mathbf{o}}^{v}$ and $B_{\mathbf{o}}^{c}=-B_{\mathbf{o}%
}^{v}$. Moreover, the stronger $V_{ME}$ leads to greater changes in the
spatial distributions of the wave functions. The spatial distributions of
the wave functions strongly depend on $k_{y}$. As for the band-edge states,
the aforementioned relationships of the wave functions are absent when $%
n^{c,v}$'s are sufficiently large enough. For small $n^{c,v}$'s (Figs.
10.18(i)-(l)), wave functions are less influenced by the modulated electric
potential. However, the spatial distribution of LS with a larger $n^{c,v}$
becomes wider (narrower) for the $k_{be}^{\varrho }$ ($k_{be}^{\varkappa }$)
state, as shown in Figs. 10.18(m)-(p). Moreover, the localization center of
the band-edge states is hardly affected by $V_{ME}(x)$.

\begin{figure}[tbp]
\par
\begin{center}
\leavevmode
\includegraphics[width=0.8\linewidth]{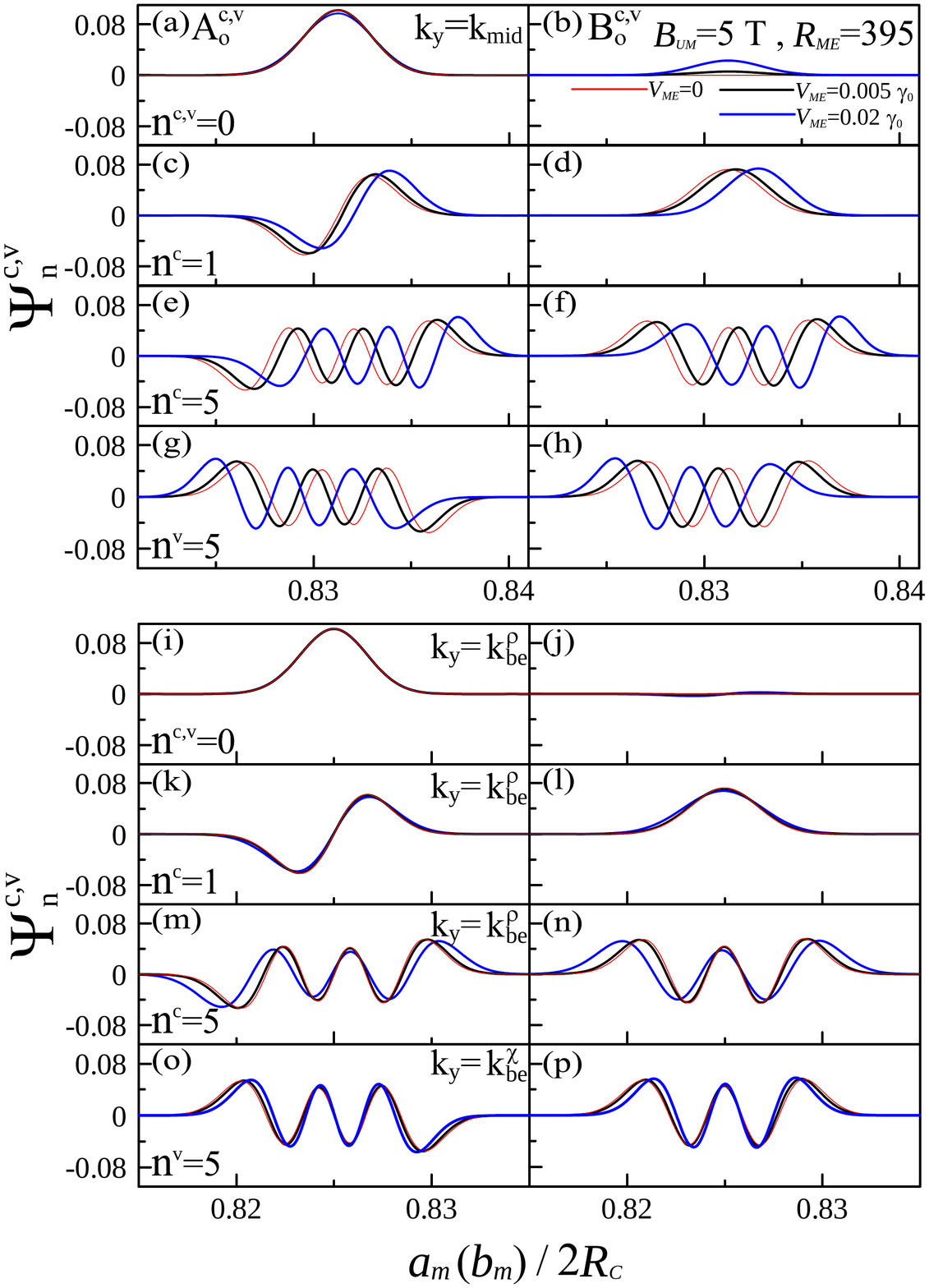}
\end{center}
\par
\textbf{Figure 10.18.} The wave functions for (a)-(d) $n^{c,v}=0$, $n^{c}=1$, $n^{c}=5$,
and $n^{v}=5$ at $k_{mid}$, (i)-(n) $n^{c,v}=0$, $n^{c}=1$ and $n^{c}=5$ at $%
k_{be}^{\rho }$, and (l)-(m) $n^{v}=5$ at $k_{be}^{\chi }$. The results
corresponding to the uniform magnetic field $B_{UM}=5$ T, the composite
field $B_{UM}=5$ T combined with $R_{MM}=395$ and $V_{ME}=0.005\gamma _{0}$,
and $B_{UM}=5$ T combined with $R_{MM}=395$ and $V_{ME}=0.02\gamma _{0}$ are
indicated by the red, black, and blue curves respectively.
\end{figure}

\subsection{Magneto-Optical Absorption Spectra Destroyed by Modulated
Electric Potentials}

Under a modulated electric potential, the changes in the electronic
properties of LLs are manifested in the optical absorption spectra. Each LL
is split into two kinds of sinusoidal subbands, with one leading the other
by $1/6$ of a period (Fig. 10.17(a)). The spatial localization region of the
wave function is completely different between two kinds of subbands (not
shown), but identical in corresponding to the same kind of LSs with
different quantum numbers at the same $k_{y}$. This indicates that the
optical transition between two different kinds of subbands is forbidden. The
absorption spectrum for $R_{ME}=395$ at $V_{ME}=0.005\gamma _{0}$ is shown
in Fig. 10.19(a) by the black line. Each absorption peak, $\omega
^{nn^{\prime }}$, originates from a transition between two LSs with quantum
numbers $n$ and $n^{\prime }$ for the same kind of subbands. In addition to
the original peaks, which correspond to the selection rule $\Delta n=1$
similar to that of LLs, there are extra peaks not characterized by the same
selection rule. In the frequency range $\omega <0.1$ $\gamma _{0}$, the peak
intensity of $\omega _{LL}^{nn+1}$, significantly reduced by $V_{ME}$,
declines as the frequency increases. The extra peaks with $\Delta n\neq 1$\
behave the opposite way. For a small $V_{ME}$, the peaks with $\Delta n=1$
are much stronger than those with $\Delta n\neq 1$.

\begin{figure}[tbp]
\par
\begin{center}
\leavevmode
\includegraphics[width=0.8\linewidth]{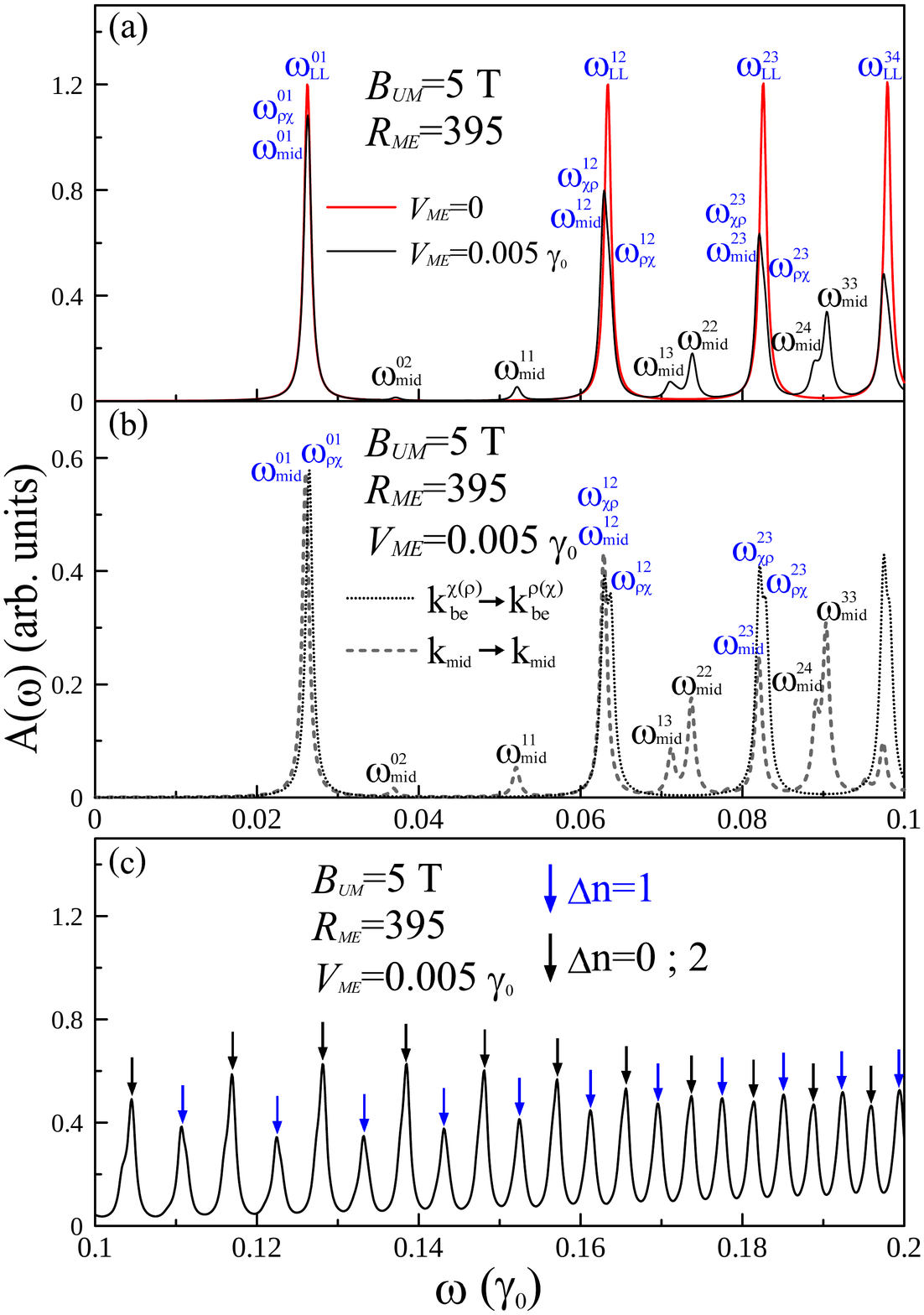}
\end{center}
\par
\textbf{Figure 10.19.} The optical absorption spectra corresponding to (a) the uniform
magnetic field $B_{UM}=5$ T by the red curve and the composite field $%
B_{UM}=5$ T combined with $R_{MM}=395$ and $V_{ME}=0.005\gamma _{0}$ by the
black curve, (b) the transitions of $k_{be}^{\chi (\rho )}\rightarrow
k_{be}^{\rho (\chi )}$ and $k_{mid}\rightarrow k_{mid}$ for $B_{UM}=5$ T
combined with $R_{MM}=395$ and $V_{ME}=0.005\gamma _{0}$ by the dotted and
dashed curves respectively, and (c) the higher frequency region of $0.1%
\symbol{126}0.2\gamma _{0}$.
\end{figure}

The original and extra peaks can be explained by the subband transitions
associated with a certain set of $k_{y}$ points. For the $k_{be}^{\varkappa
(\varrho )}\rightarrow k_{be}^{\varrho (\varkappa )}$ transitions,\ the
corresponding band-edge states have a high DOS and the symmetry of wave
function is little changed. Therefore, they can promote the prominent peaks
consistent with the $\Delta n=1$ selection rule.\ As shown in Fig. 10.19(b)
by the thin dashed lines, $\omega _{\varkappa \varrho }^{nn+1}$\ and\ $%
\omega _{\varrho \varkappa }^{nn+1}$ represent the absorption peaks from
the\ transition channels [$nk_{be}^{\varkappa }\rightarrow $($n+1$)$%
k_{be}^{\varrho }$, ($n+1$)$k_{be}^{\varrho }\rightarrow nk_{be}^{\varkappa
} $] and [$nk_{be}^{\varrho }\rightarrow $($n+1$)$k_{be}^{\varkappa }$, ($%
n+1 $)$k_{be}^{\varkappa }\rightarrow nk_{be}^{\varrho }$], respectively.
These two peaks are close to each other, and almost overlap with the
original peaks. But in cases where either $n$ or $n^{_{\prime }}$ is zero,
only the peak $\omega _{\varrho \varkappa }^{01}$ can be created by two
transition channels: [$0k_{be}^{\varrho }\rightarrow 1k_{be}^{\varkappa }$, $%
1k_{be}^{\varkappa }\rightarrow 0k_{be}^{\varrho }$]. The reduction of the
transition channels is due to the fact that the $n^{c,v}=0$ subbands
oscillate between the conduction and valence bands. For the $%
k_{mid}\rightarrow k_{mid}$ transitions, the significant change to the
symmetry of the wave function results in different selection rules, i.e., $%
\Delta n\neq 1$. As shown in Fig. 10.19(b) by the thick dashed lines, the
absorption peak $\omega _{mid}^{nn^{\prime }}$ corresponds to the
transitions between two LSs of $n^{v(c)}$ and $n^{_{\prime }c(v)}$ from the $%
k_{mid}$ states. These types of peaks are responsible for the extra peaks in
Fig. 10.19(a), i.e., the $\Delta n\neq 1$ peaks primarily arise from the
middle states at lower modulation strength. The $k_{mid}\rightarrow k_{mid}$
transitions also contribute to the original peaks. Such contributions
decline with an increased frequency. However, the $\Delta n=1$ absorption
peaks are dominated by both the band-edge and middle states.

The peak intensities from different selection rules change dramatically with
respect to the variation in frequency. The intensities of the $\Delta n=1$
peaks gradually rise for a further increase in frequency, while the opposite
is true for those of the $\Delta n\neq 1$ peaks (Fig. 10.19(c)). Within the
frequency range of $0.1$ $\gamma _{0}<\omega <0.2$ $\gamma _{0}$, the former
is lower than the latter. The main reason for this is the fact that the
symmetry violation of the wave functions is enhanced for LSs with larger $%
n^{c,v}$'s. It is deduced that the extra absorption peaks of $\Delta n\neq 1$
are relatively easily observed for experimental measurements at higher
frequencies. Each of them is composed of two peaks, $\omega _{mid}^{nn}$ and
$\omega _{mid}^{n-1n+1}$, which satisfy $\Delta n=0$ and $\Delta n=2$ with
nearly the same frequency.

The absorption spectrum exhibits more features for a higher modulated
potential, as shown in Fig. 10.20 for $R_{ME}=395$ at $V_{ME}=0$, $%
0.02\gamma _{0}$. In Fig. 10.20(a), the intensity of the extra peaks becomes
comparable to that of the original peaks. Unlike in the lower $V_{ME}$ case
(Fig. 10.19(a)), the peak intensities, regardless of their types, vary
irregularly with the frequency. The oscillations of the subband structure
are obviously augmented and cause the subband transitions to change greatly
with respect to $k_{y}$. As a result, the absorption peaks grow much wider
and split more evenly. Peaks generated by different selection rules are
likely to appear, owing to the severe breakdown of the spatial symmetry of
the wave functions. It should be noted that it is difficult to distinguish
the original from the extra peaks solely based on the peak heights. Besides
the aforementioned spectrum analysis, further discussions are made regarding
two specific $k_{y}$ points. For the $k_{be}^{\varkappa (\varrho
)}\rightarrow k_{be}^{\varrho (\varkappa )}$ transitions, $\varkappa $- and $%
\varrho $-type band-edge states show different behavior in terms of energy
spacing and curvature such that the two transition channels [$%
nk_{be}^{\varkappa }\rightarrow $($n+1$)$k_{be}^{\varrho }$, ($n+1$)$%
k_{be}^{\varrho }\rightarrow nk_{be}^{\varkappa }$] and [$nk_{be}^{\varrho
}\rightarrow $($n+1$)$k_{be}^{\varkappa }$, ($n+1$)$k_{be}^{\varkappa
}\rightarrow nk_{be}^{\varrho }$] possess distinct frequencies (thin dashed
line in Fig. 10.20(b)). Therefore, the initially coinciding peaks split up,
and the original intensities are divided into fractions. Moreover, the
band-edge states can induce the extra peaks of $\Delta n=0$, e.g., $\omega
_{\varrho \varkappa }^{11}$ and $\omega _{\varrho \varkappa }^{22}$. When $%
k_{y}=k_{mid}$, the symmetry of the wave functions is destroyed.
Consequently, the extra peaks are strengthened, and the original peaks are
weakened or even disappear (thick dashed line in Fig. 10.20(b)). The
corruption of the orthogonality of the sublattices $A_{\mathbf{o}}^{c(v)}$
and $B_{\mathbf{o}}^{v(c)}$ enables the subband transitions to occur from $%
\Delta n=3$; such examples are seen in $\omega _{mid}^{03}$ and $\omega
_{mid}^{14}$. The very strong dispersions of LSs also lead to the splitting
of the original peaks associated with $k_{mid}$. This can account for the ($%
\omega _{mid}^{01}$, $\omega _{\varrho \varkappa }^{01}$) peaks and the ($%
\omega _{mid}^{23}$, $\omega _{\varkappa \varrho }^{23}$, $\omega _{\varrho
\varkappa }^{23}$) peaks in Fig. 10.20(a).

\begin{figure}[tbp]
\par
\begin{center}
\leavevmode
\includegraphics[width=0.8\linewidth]{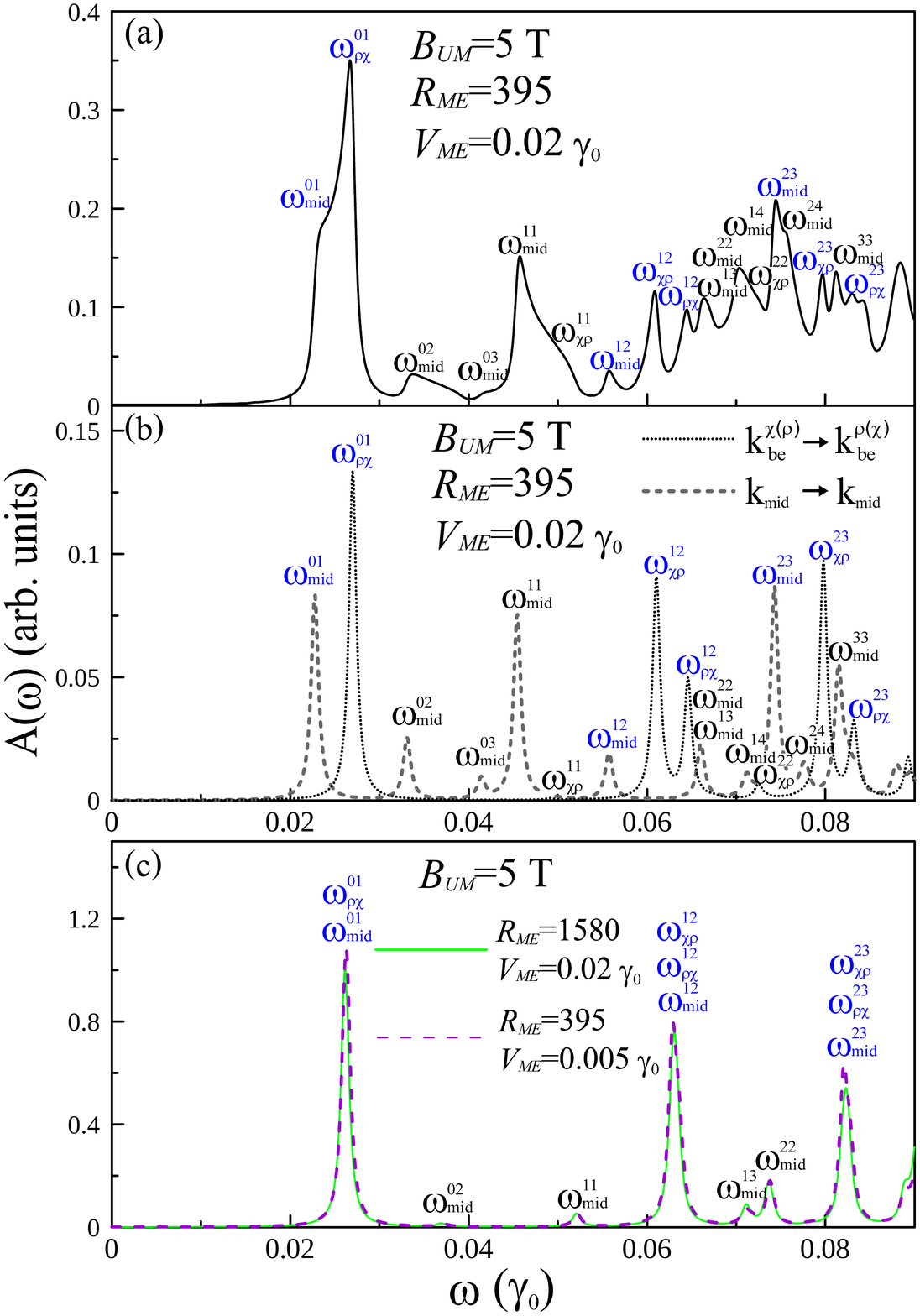}
\end{center}
\par
\textbf{Figure 10.20.} The optical absorption spectra for the composite field $%
B_{UM}=5$ T combined with $R_{MM}=395$ and $V_{ME}=0.02\gamma _{0}$, where
the contributions from the transitions $k_{be}^{\chi (\rho )}\rightarrow
k_{be}^{\rho (\chi )}$ and $k_{mid}\rightarrow k_{mid}$ are shown in (b) by
the dotted and dashed curves respectively. (c) A comparison between the
absorption spectra for $B_{UM}=5$ T combined with $R_{MM}=395$ and $%
V_{ME}=0.005\gamma _{0}$ and $B_{UM}=5$ T combined with $R_{MM}=1580$ and $%
V_{ME}=0.02\gamma _{0}$.
\end{figure}

The modulation period strongly affects the magneto-optical spectrum of the
LSs, while the modulation effect is less significant with sufficiently
larger period. The absorption peaks from $\Delta n=1$ are much higher than
those from $\Delta n\neq 1$, as shown by the green line in Fig. 10.20(c) for
$V_{ME}=0.02$ $\gamma _{0}$ at a larger period $R=1580$. The transition
energies between the valence and conduction LSs at different $k_{y}$ values
are almost the same. Thus, the frequencies $\omega _{\varkappa \varrho
}^{nn+1}$, $\omega _{\varrho \varkappa }^{nn+1}$, and $\omega _{mid}^{nn+1}$
associated with these LS transitions are all the same. Since the symmetry of
the wave function does not degrade much, the absorption peaks of $\Delta n=3$
no longer exist. The modulated electric field $\mathbf{E}=-\mathbf{\nabla }%
_{x}V_{ME}(x)=(2\pi V_{ME}/3b^{\prime }R_{ME})\sin (2\pi x/3b^{\prime
}R_{ME})$$\widehat{x}$ implies that the same value of $V_{ME}/R_{ME}$
produces the same modulation effect. For instance, the absorption spectrum
for $R_{ME}=395$ at $V_{ME}=0.005\gamma _{0}$ (the purple dashed curve in
Fig. 10.20(c)) is almost same as that for $R_{ME}=395$ at $V_{ME}=0$, $%
0.005\gamma _{0}$ owing to $V_{ME}/R_{ME}=$ $0.005/395=0.02/1580$.

We look at the relationship between the absorption frequency and the
potential strength more closely. At $V_{ME}=0$, the peaks denoted by the
blue symbols in Fig. 10.21 can only appear if they obey the selection rule $%
\Delta n=1$. After an external modulation potential is applied, peaks from
other selection rules ($\Delta n=0$ and $\Delta n=2$) appear as shown by the
black symbols. Under a weak modulated potential ($V_{ME}<0.005$ $\gamma _{0}$%
), the peak frequency is hardly affected. If the electric potential
continues to grow, some peaks start to split, e.g., $\omega _{\varkappa
\varrho }^{nn+1}$, $\omega _{\varrho \varkappa }^{nn+1}$ and $\omega
_{mid}^{nn+1}$. Under a strong potential ($V_{ME}>0.02$ $\gamma _{0}$), more
peaks are created by other selection rules, such as $\Delta n=3$ (square
markers). Even for\ the transitions between band-edge states, peaks can be
developed by the rule $\Delta n\neq 1$ (solid triangles). Finally, the
absorption frequencies related to the $k_{mid}$ states\ decrease rapidly
with respect to the increment of $V_{ME}$. This is due to the fact that the
conduction and valence LSs at $k_{mid}$ move closer to the Fermi level.
These theoretical predictions could be examined by the optical-absorption
spectroscopy methods.\cite{10.2.7,cpc42,cpc43,cpc44,cpc45}

\begin{figure}[tbp]
\par
\begin{center}
\leavevmode
\includegraphics[width=0.8\linewidth]{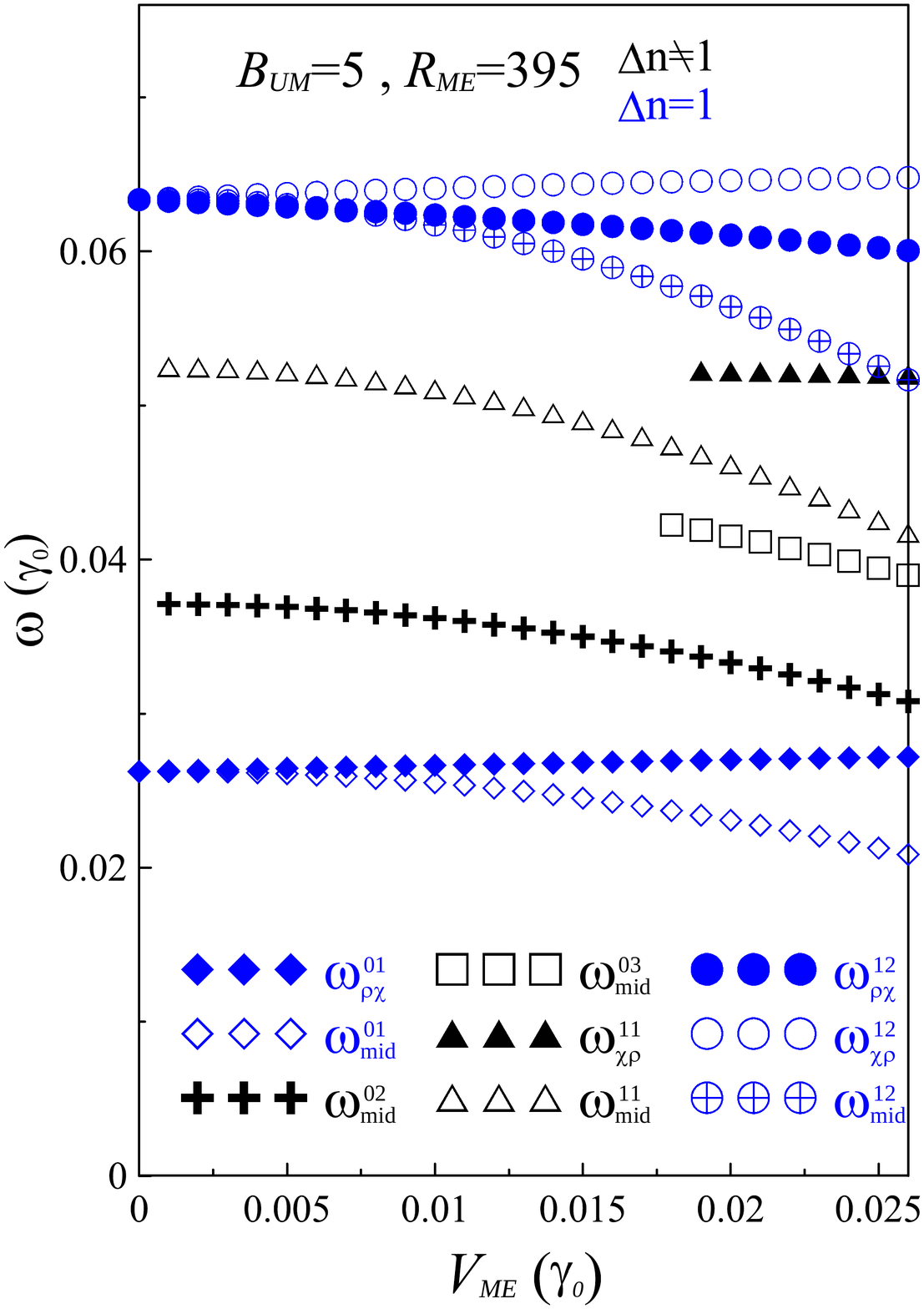}
\end{center}
\par
\textbf{Figure 10.21.} The dependence of absorption frequencies, $\omega _{\chi \rho
}^{nn^{\prime }}$, $\omega _{\rho \chi }^{nn^{\prime }}$, and $\omega
_{mid}^{nn^{\prime }}$, on the modulated strength $V_{ME}$. The absorption
frequencies with the selection rules $\left\vert \Delta n\right\vert
=\left\vert n-n^{\prime }\right\vert =1$ and $\left\vert \Delta n\right\vert
\neq 1$ are indicated by the blue and black colors respectively.
\end{figure}

\newpage

\section{Conclusion}

The results show that monolayer graphene exhibits the rich optical
absorption spectra, an effect being controlled by the external fields. Such
fields have a strong influence on the number, intensity, frequency and
structure of absorption peaks. Moreover, there would exist the dissimilar
selection rules for different external fields. In the presence of the
uniform magnetic field, the magneto-optical excitations obey the specific
selection rule $\Delta n=1$, since the simple relationship exists between
the two sublattices of $a$- and $b$-atoms. As to a modulated magnetic field,
an extra selection rule $\Delta n=0$ is obtained, due to the complex
overlapping behavior from two subenvelope functions in the wave function.
However, the wave functions exhibit irregular behaviors under the modulated
electric potential. As a result, it is difficult to single out a particular
selection rule. In the composite fields, the symmetry of the LL wave
functions is broken by the introduce of two kinds of modulation fields, the
modulated magnetic and electric fields, a cause resulting in the altered
selection rules in the absorption excitations. The one case is the uniform
magnetic field combined with the modulated magnetic field. The extra
selection rules, e.g., $\Delta n=2$ and $3$, come to exist when $B_{MM}$ is
comparable to $B_{UM}$. Another case is the uniform magnetic field combined
with the modulated electric potential. The extra selection rules, e.g., $%
\Delta n=0$, $2$ and $3$, would be generated in the increase of $V_{ME}$.

For the other graphene systems, the magneto-optical properties corresponding
to a perpendicular uniform magnetic field deserve a closer investigation.
For example, the AA-stacked bilayer graphene is predicted to exhibit two
groups of absorption peaks;\cite{10.1.4} however, the selection rule $\Delta
n=1$ is same as that of MG. As for the AB-stacked bilayer graphene, there
exist four groups of absorption peaks and two extra selection rules ($\Delta
n=0$ and $2$). The few-layer graphenes are expected to display more complex
magneto-optical absorption spectra, mainly owing to the number of layers and
the stacking configuration.

In this chapter, the generalized tight-binding model is introduced to
discuss monolayer graphene under five kinds of external fields. The
Hamiltonian, which determines the magneto-electronic properties, is a giant
Hermitian matrix for the experimental fields. It is transformed into a
band-like matrix by rearranging the tight-binding functions; furthermore,
the characteristics of wave function distributions in the sublattices are
used to reduce the numerical computation time. In the generalized
tight-binding model, the $\pi $-electronic structure of MG is solved in the
wide energy range of $\pm 6$ eV, a solution proving valid even if the
magnetic, electric or composite field is applied. Moreover, the important
interlayer atomic interactions, not just treated as the perturbations, could
be simultaneous included in the calculations. The generalized model can also
be extensible to other layer stacked systems, i.e., AA-, AB-, ABC-stacked
FLGs \cite{p224,p223,p220,p216,p209} and bulk graphite.\cite%
{RBChen,p228,p215,p208}

\bigskip \vskip0.6 truecm\newpage

\centerline {\Large \textbf {Figure Captions}}

\vskip0.5 truecm

Fig. 10.1 The primitive unit cell of monolayer graphene (a) in the absence
and (b) in the presence of external fields.

\vskip0.5 truecm

Fig. 10.2 (a) The Landau level spectrum for the uniform magnetic field $%
B_{UM}=5$ T. The Landau level wave functions corresponding to (b) the $a$-
and (c) the $b$-atoms.

\vskip0.5 truecm

Fig. 10.3 The optical absorption spectra for (a) $B_{UM}=5$ T and (b) $%
B_{UM}=10$ T. (c) The dependence of the absorption frequency on the square
root of field strength $B_{UM}$.

\vskip0.5 truecm

Fig. 10.4 The energy dispersions and the illustration of optical excitation
channels for the modulated magnetic field along the armchair direction with $%
R_{MM}=500$ and $B_{MM}=10$ T.

\vskip0.5 truecm

Fig. 10.5 The wave functions of Quasi-Landau levels at (a)-(f) the original
band-edge state $k_{1}$ with the quantum numbers $n^{c,v}=0$, $n^{c}=1$ and $%
n^{c}=2$, (g) and (h) the split point $k_{2}$ with $n^{c}=1$, and (i)-(l)
two extra band-edge states $k_{3}$ and $k_{4}$ with $n^{c}=1$.

\vskip0.5 truecm

Fig. 10.6 The optical absorption spectra for (a) $B_{MM}=10$ T at a fixed
periodic length with modulation and polarization along the armchair and
zigzag directions and (b) different modulation periods and field strengths
with both the modulation and polarization along the armchair direction.

\vskip0.5 truecm

Fig. 10.7 The dependence of the absorption frequency on (a) the period $%
R_{MM}$ and (b) the square root of field strength $B_{MM}$.

\vskip0.5 truecm

Fig. 10.8 The energy dispersions for the modulated electric potential along
the armchair direction with $R_{MM}=500$ and $V_{ME}=0.05\gamma _{0}$.

\vskip0.5 truecm

Fig. 10.9 The optical absorption spectra corresponding to Fig. 10.8, which
includes the contributions from the $\mu $ and $\nu $ states, respectively.

\vskip0.5 truecm

Fig. 10.10 The optical absorption for (a) $V_{ME}=0.05\gamma _{0}$ at a
fixed periodic length with modulation and polarization along the armchair
and zigzag directions and (b) different modulation periods and field
strengths with both the modulation and polarization along the armchair
direction.

\vskip0.5 truecm

Fig. 10.11 The wave functions at different band-edge states, the $\mu
_{i}^{c}$ and $\nu _{i}^{c}$ states for $i=1\symbol{126}7$.

\vskip0.5 truecm

Fig. 10.12 The energy dispersions for (a) the uniform magnetic field $%
B_{UM}=5$ T by the red curves and the composite field $B_{UM}=5$ T combined
with $R_{MM}=395$ and $B_{MM}=1$ T by the black curves, (b) $B_{UM}=5$ T
combined with $R_{MM}=395$ and $B_{MM}=5$ T, and (c) $B_{UM}=5$ T combined
with $R_{MM}=395$ and $B_{MM}=40$ T. All modulated fields are applied along
the armchair direction.

\vskip0.5 truecm

Fig. 10.13 The wave functions with $n^{c,v}=0$ and $n^{c}=1$ at $%
k_{be}^{\zeta }$ for (a)-(d) the uniform magnetic field $B_{UM}=5$ T by the
red curves and the composite field $B_{UM}=5$ T combined with $R_{MM}=395$
and $B_{MM}=1$ T by the black curves, (e)-(h) $B_{UM}=5$ T by the red curves
and $B_{UM}=5$ T combined with $R_{MM}=395$ and $B_{MM}=5$ T by the black
curves, and (i)-(l) $B_{UM}=5$ T combined with $R_{MM}=395$ and $B_{MM}=40$
T.

\vskip0.5 truecm

Fig. 10.14 The optical absorption spectra corresponding to (a) the uniform
magnetic field $B_{UM}=5$ T by the red curve and the composite field $%
B_{UM}=5$ T combined with $R_{MM}=395$ and $B_{MM}=1$ T by the black curve
and (b) the composite field $B_{UM}=5$ T combined with $R_{MM}=395$ and $%
B_{MM}=5$ T.

\vskip0.5 truecm

Fig. 10.15 The optical absorption spectra corresponding to the composite
field $B_{UM}=5$ T combined with $R_{MM}=395$ and $B_{MM}=40$ T by the red
dotted curve and the pure modulated magnetic field $R_{MM}=395$ and $%
B_{MM}=40$ T by the black curve.

\vskip0.5 truecm

Fig. 10.16 The dependence of absorption frequencies, $\omega _{\eta
}^{nn^{\prime }}$ and $\omega _{\zeta }^{nn^{\prime }}$ with $\left\vert
\Delta n\right\vert =\left\vert n-n^{\prime }\right\vert =1$, on the
modulated strength $B_{MM}$.

\vskip0.5 truecm

Fig. 10.17 The energy dispersions for (a) the uniform magnetic field $%
B_{UM}=5$ T by the red curves and the composite field $B_{UM}=5$ T combined
with $R_{MM}=395$ and $V_{ME}=0.005\gamma _{0}$ by the black curves, (b) $%
B_{UM}=5$ T combined with $R_{MM}=395$ and $V_{ME}=0.02\gamma _{0}$. All
modulated fields are applied along the armchair direction.

\vskip0.5 truecm

Fig. 10.18 The wave functions for (a)-(d) $n^{c,v}=0$, $n^{c}=1$, $n^{c}=5$,
and $n^{v}=5$ at $k_{mid}$, (i)-(n) $n^{c,v}=0$, $n^{c}=1$ and $n^{c}=5$ at $%
k_{be}^{\rho }$, and (l)-(m) $n^{v}=5$ at $k_{be}^{\chi }$. The results
corresponding to the uniform magnetic field $B_{UM}=5$ T, the composite
field $B_{UM}=5$ T combined with $R_{MM}=395$ and $V_{ME}=0.005\gamma _{0}$,
and $B_{UM}=5$ T combined with $R_{MM}=395$ and $V_{ME}=0.02\gamma _{0}$ are
indicated by the red, black, and blue curves respectively.

\vskip0.5 truecm

Fig. 10.19 The optical absorption spectra corresponding to (a) the uniform
magnetic field $B_{UM}=5$ T by the red curve and the composite field $%
B_{UM}=5$ T combined with $R_{MM}=395$ and $V_{ME}=0.005\gamma _{0}$ by the
black curve, (b) the transitions of $k_{be}^{\chi (\rho )}\rightarrow
k_{be}^{\rho (\chi )}$ and $k_{mid}\rightarrow k_{mid}$ for $B_{UM}=5$ T
combined with $R_{MM}=395$ and $V_{ME}=0.005\gamma _{0}$ by the dotted and
dashed curves respectively, and (c) the higher frequency region of $0.1%
\symbol{126}0.2\gamma _{0}$.

\vskip0.5 truecm

Fig. 10.20 (a) The optical absorption spectra for the composite field $%
B_{UM}=5$ T combined with $R_{MM}=395$ and $V_{ME}=0.02\gamma _{0}$, where
the contributions from the transitions $k_{be}^{\chi (\rho )}\rightarrow
k_{be}^{\rho (\chi )}$ and $k_{mid}\rightarrow k_{mid}$ are shown in (b) by
the dotted and dashed curves respectively. (c) A comparison between the
absorption spectra for $B_{UM}=5$ T combined with $R_{MM}=395$ and $%
V_{ME}=0.005\gamma _{0}$ and $B_{UM}=5$ T combined with $R_{MM}=1580$ and $%
V_{ME}=0.02\gamma _{0}$.

\vskip0.5 truecm

Fig. 10.21 The dependence of absorption frequencies, $\omega _{\chi \rho
}^{nn^{\prime }}$, $\omega _{\rho \chi }^{nn^{\prime }}$, and $\omega
_{mid}^{nn^{\prime }}$, on the modulated strength $V_{ME}$. The absorption
frequencies with the selection rules $\left\vert \Delta n\right\vert
=\left\vert n-n^{\prime }\right\vert =1$ and $\left\vert \Delta n\right\vert
\neq 1$ are indicated by the blue and black colors respectively.
\end{subequations}

\end{document}